# On the Linear Stability of Sheared and Magnetized Jets Without Current Sheets – Non-Relativistic Case

By

Jinho Kim[1]⋆, Dinshaw S. Balsara[1]⋆, Maxim Lyutikov[2], Serguei S. Komissarov[3]

[1]*Physics Department, College of Science, University of Notre Dame, 225 Nieuwland Science Hall, Notre Dame, IN 46556, USA*

[2]*Department of Physics, Purdue University, 525 Northwestern Avenue, West Lafayette, IN 47907-2036, USA*

[3]*Department of Applied Mathematics, The University of Leeds, Leeds LS2 9GT*

**Abstract**

In a prior paper (Kim *et al*. 2015) we considered the linear stability of magnetized jets that carry no net electric current and do not have current sheets. In this paper, in addition to physically well-motivated magnetic field structures, we also include the effects of jet shear. The jets we study have finite thermal pressure in addition to having realistic magnetic field structures and velocity shear.

We find that shear has a strongly stabilizing effect on various modes of jet instability. Increasing shear stabilizes the fundamental pinch modes at long wavelengths and short wavelengths. Increasing shear also stabilizes the first reflection pinch modes at short wavelengths. Increasing shear has only a very modest stabilizing effect on the fundamental kink modes at long wavelengths; however, increasing shear does have a strong stabilizing effect on the fundamental kink modes at short wavelengths. The first reflection kink modes are strongly stabilized by increasing shear at shorter wavelengths. Overall, we find that the combined effect of magnetic field and shear stabilizes jets more than shear alone. In addition to the results from a formal linear stability analysis, we present a novel way of visualizing and understanding jet stability. This gives us a deeper understanding of the enhanced stability of sheared, magnetized jets.

We also emphasize the value of our numerical approach in understanding the linear stability of jets with realistic structure.

⋆ E-mail: jkim46@nd.edu (JK); dbalsara@nd.edu (DSB)

# I) Introduction

Jets arise quite frequently in energetic astrophysical systems. Extragalactic jets emerging from active galactic nuclei (AGN; Rees 1978) can span several tens of Kpc to several Mpc. Equally spectacular jets emerge from young stars (Reipurth et al. 1988) and these jets too are known to propagate several parsecs from their source in young stellar objects (YSOs). X-ray binaries and gamma-ray bursters are also known to be sources of jet activity. While an observational elucidation of jet acceleration mechanisms is lacking, theorists agree that strong magnetic fields play a role in jet formation and acceleration (e.g. Lovelace 1976; Blandford & Znajek 1977; Blandford & Payne 1982; Komissarov & McKinney 2007; McKinney & Narayan 2007; Komissarov & Barkov 2009; McKinney & Blandford 2009). Observations of extragalactic jets do not permit a direct measurement of the jet's magnetic field, as a result, one has to resort to an equipartition hypothesis. This is based on an assumption that the jet's magnetic pressure is in equipartition with its thermal pressure.

Observations of astrophysical jets indicate that they propagate with unusually good stability. I.e. the jets manage to retain their structural integrity over huge distances compared to their initial radii. For example, jets from young stars should emerge on scales that are comparable to the magnetospheres of YSOs (Ray 2012), yet they propagate out to distances that can be $\sim 10^5$ to $10^7$ times larger than their natal radii. For AGN jets this ratio becomes even larger, reaching $10^9$. Comparing these ratios to the propagation lengths of terrestrial jets, we find that terrestrial jets usually destabilize over distances of tens to hundreds of jet radii. Plasma experiments have shown that magnetic fields may help in stabilizing jets; however, magnetic fields can also be the source of further instabilities. For this reason, the present paper looks to shear in the jets' channel as a source of stabilization for magnetized astrophysical jets.

The traditional method for studying the stability of astrophysical jets has been linear stability analysis. Simpler stability analyses result in linearized systems of equations with analytical solutions. Usually, these solutions turn out to be Bessel functions or hypergeometric functions. This quest for analytical tractability is desirable because it permits many examples of jets to be explored with minimal computational resources and minimal computer coding. Unfortunately, making this choice also forces the theorist to make further compromises. For example, one may choose simpler top-hat velocity profiles for the jet. Likewise, simple structures are sought for the magnetic field. Attempts to consider non-trivial magnetic field profiles within the jets have usually forced theorists to consider pressure-free jets. However, thermal effects are known to produce an isotropic pressure profile which has a significant stabilizing role in plasmas. Going beyond these simplifying assumptions required breaking free of a dependence on analytical solutions. I.e., the linear systems that arise from stability analysis have to be solved numerically. This requires a deeper investment in computer codes and computational resources. A fully numerical approach to stability analysis was first attempted by us in Kim *et al*. (2015) who studied jets with very special magnetic field properties suggested by Gourgouliatos *et al*. (2012). While our prior study included very sophisticated magnetic field topologies, the velocity distribution was still restricted to a top-hat profile. In this paper, we go past that restriction.

As seen from the previous paragraphs, prior authors have studied jet stability with certain limiting approximations. For example, some authors have only considered top-hat velocity profiles (Hardee 1979, 1982; Cohn 1983; Payne & Cohn 1985; Istomin & Pariev 1996; Begelman 1998; Lyubarskii 1999). Other authors have studied simpler magnetic field structures, restricting attention to either axial or toroidal magnetic fields in the jets (Istomin & Pariev 1994, 1996; Narayan et al. 2009). Some recent work has tried to include more complex magnetic field geometries (Bodo *et al.* 2013) but that frequently comes at the expense of restricting the study to a consideration to pressure-free jets, which is not realistic. In Kim *et al.* (2015) we started a line of inquiry which removed many of these simplifying approximations. For example, the jets that were studied in our prior paper were quite realistic because they included complex magnetic field geometries while retaining finite thermal pressure in the jets. Realistic jets also could very likely have sheared velocity profiles. The *first goal* of this paper is to study the stability of jets with realistic magnetic field structure, realistic pressure support and realistic velocity shear. We realize, therefore, the perturbed jets that emerge from our stability analysis will have very complicated perturbation structure. The *second goal* of this paper is, therefore, to present a novel way of visualizing and understanding jet stability in the linear regime. While prior stability analyses were entirely analytical and based on the structure of the Bessel function, the work presented here is mostly numerical. The *third goal* of this paper is to emphasize the value of this numerical approach in understanding the linear stability of jets with realistic structure.

In this paper we present a novel way of visualizing and understanding jet stability. In our approach, it is better to focus on a given family of modes. The mode families that we are interested in are the fundamental mode of the pinch instability, the first reflection mode of the pinch instability, the fundamental mode of the kink instability and the first reflection mode of the kink instability. As a result, each of those modes is given its own Section and is studied independently. Section II describes the unperturbed jet structure of the sheared, magnetized jets without surficial current sheets. It also mentions our stability analysis methods for the sake of completeness. Section III describes the stability of the fundamental mode of the pinch instability and shows that it is possible to arrive at a deeper understanding its the stability via the visual methods developed here. Section IV does the same for the first reflection mode of the pinch instability. Section V presents a study of the fundamental mode of the kink instability and Section VI focuses on the first reflection mode of the kink instability. Section VII provides discussion and conclusions.

**II) Description of the Unperturbed Jet Structure and Our Stability Analysis**

This paper is a sequel to Kim *et al.* (2015). Consequently, all the notation is kept entirely consistent from the previous paper to this paper. It is very important to catalogue the structure of the jets that we consider here. All the magnetized jets that we consider here have zero net electric current in them. This is achieved by having a special form of the magnetic field, as shown by Gourgouliatos *et al.* (2012). The special form of magnetic field effectively replaces surface currents with a current density that is distributed across the volume of the jet. Fig. 1, which is drawn from Gourgouliatos *et al.* (2012), shows the axial and toroidal magnetic field in the jet as a function of radius. Notice that the magnetic field is purely axial on-axis; however, the toroidal magnetic field becomes more dominant as one moves outward from the axis. This form of magnetic field was initially inspired by tokamak and spheromak studies and has a net zero magnetic field at the boundary of the jet. The end result is that, unlike prior studies on the stability of magnetized jets, there is no concentrated current sheet at the surface of the jets that we consider here. While we do not consider resistive effects in this paper, one of the beneficial consequences of the magnetic field profile that is used in our present study is that there are no concentrated resistive instabilities at the boundaries of the jets. The absence of concentrated current sheets at the boundaries of the jets might also have positive consequences for numerical simulations. In this paper we find that the shear in the jets' axial velocity also plays an extremely important role in stabilizing the jets. For that reason, it is important to be able to parametrize the jets' axial velocity. We do this by specifying the jets' velocity with the formula

$$v_{z0}(r) = v_{z;\max}\left(1 - a\left(\frac{r}{r_j}\right)^2\right)$$

In the above formula, $v_{z;\max}$ is the maximal on-axis velocity of the jets, $r_j$ is the jets' radius and is usually set to unity. The parameter "$a$" ranges from 0 for a top-hat velocity profile to 0.9 for a jet whose axial velocity almost blends in with the ambient velocity. Fig. 2 shows the four velocity profiles considered in this paper with $a = 0, 0.3, 0.6, 0.9$.

Please note that Rayleigh (1896) had found that such parabolic profiles are stable for incompressible couette flow and that is our intuitive motivation for thinking that sheared jets might have enhanced stability. It is quite possible that even a jet that starts off with a top-hat profile might entrain ambient material through its boundary, thus reaching a parabolic velocity profile. Furthermore, jet launching mechanisms need to tie into the magnetic field structure at the central black hole, thus giving rise to a more sheared profile.

The jets that we consider are all pressure-matched with their ambient medium. The ambient medium is uniform and unmagnetized, so that the thermal pressure in the ambient medium matches the total (gas + magnetic) pressure at the surface of the jet. All linear stability analyses are based on jets with a constant entropy in the unperturbed jet. In keeping with that trend, we consider the entropy in the jet fluid to be a constant. Unlike most prior stability analyses, the magnetic field in our jets has a non-trivial structure, see Fig. 1. The total unperturbed pressure at the jet boundary must balance the gas pressure in the ambient medium. However, Fig. 1 shows that the jet's magnetic pressure varies with jet radius. As a result, force balance in the radial direction requires the gas pressure to also vary with jet radius. The varying gas pressure, along with the constant entropy requirement, also requires the density to vary as a function of radius. In this paper, we will be using some of these parameters as measured at the jet axis. Thus, one parameter that catalogues different jets is given by $\eta = \rho_j / \rho_a$, where $\rho_j$ is the jet density as measured at the jet axis and $\rho_a$ is the uniform external density in the ambient medium. The on-axis gas pressure is also determined by pressure-balance considerations. The pressure, however, varies as a function of radius. In such a jet, it is important to have some way of defining the mean Mach number. We define our mean Mach number by a scaled ratio of the jet kinetic energy to the jet thermal energy. The definition given below is such that when the jet density, velocity and pressure are constant (i.e. jet with top-hat profile), the equation below reduces to the Mach number. Our definition of the jet Mach number is, therefore, given by

$$M = \sqrt{\frac{\int_0^{r_j} \rho_0(r) v_{z0}^2(r)\, r\, dr}{\gamma \int_0^{r_j} P_0(r)\, r\, dr}}$$

In the above equation, $\rho_0(r)$ is the unperturbed density profile across the jet; $v_{z0}(r)$ is the unperturbed velocity profile across the jet; $P_0(r)$ is the unperturbed pressure profile across the jet and $\gamma$ is the polytropic index. In this paper we adopt a novel way of visualizing jet stability and we will always inter-compare jets with the same Mach number, where the Mach number is defined by the above equation. For magnetized jets, the plasma-$\beta$ also plays an important role in determining jet stability. The on-axis plasma-$\beta$ was used in the previous paper (Kim *et al*. 2015) and we follow the same definition here. The extent of the shear, as parametrized by the value of "*a*" in Fig. 2, constitutes the fourth parameter specifying the jet. Thus any jet in our simulations is specified by four parameters – the density ratio $\eta$, the Mach number "*M*", the plasma-$\beta$ and the extent of the shear "*a*".

In Sections 2 and 3 of Kim *et al*. (2015) we have already described our numerically-motivated strategy for carrying out stability analyses of jets with realistic density, pressure and magnetic field profiles. As in our previous study, we study perturbations to the jet's flow variables of the form $\delta f(t,r,\phi,z) = f(r)\exp(i\omega t - im\phi - ikz)$, where *m*=0 or 1 modes, i.e. for pinch or kink modes. For our purposes, we keep the wave number "*k*" real, while allowing the angular frequency, "$\omega$", to be complex. The real part of the angular frequency "$\omega$" gives us the angular oscillation frequency of the jet, while its imaginary part gives us the growth rate. Specification of the functions $f(r)$ for each of the flow variables in the jet gives us the eigenfunction of the jet. Our formulation is very general and is brought over unchanged from our previous paper to the present paper. For that reason, we do not repeat our description of how the stability analysis was carried out in this paper, instead we refer the interested reader to our prior paper. The novel element in this paper is the inclusion of shear in the jets. However, we point out that the Sections 2 and 3 of Kim *et al*. (2015) already describe the inclusion of shear in the jet's profile even though the stability analysis in the previous paper was carried out for jets with top-hat velocity profiles.

The jets that we consider in this paper have *M*=4 and $\eta$=0.1; i.e. they are supersonic light jets. The jets have a range of values of "*a*", leading to several different values of shear. They also have a range of values of "$\beta$", leading to different amounts of magnetic pressure. The different values of shear that we consider correspond to *a*=0 (top-hat velocity profile), *a*=0.3 (mild shear), *a*=0.6 (modest shear) and *a*=0.9 (strong shear). We also explore values of "$\beta$" corresponding to $\beta = \infty$ (no magnetic field), $\beta = 1$ (equipartition magnetic field) and $\beta = 0.5$ (strong magnetic field. Taken together, this constitutes twelve different light, supersonic jets whose stability we explore here with our linear stability analysis.

Numerous studies of the stability of astrophysical jets have shown that they have at least some instability. (In fact, the presence of some instabilities might even be a good thing if the waves associated with the instabilities cause particles to be accelerated in the jet, thereby enabling the jets to shine with radio emission.) For one to say that the stability of a jet is improved by the inclusion of shear or magnetic field, one has to specify some criterion for saying that the stability is improved. Consider a jet that is perturbed with an unstable mode with wave number "$k$" having a complex frequency $\omega = \omega_R + i\omega_I$ with $\omega_I < 0$. Consequently, $\tau_I = 1/|\omega_I|$ measures the time in which the unstable jet undergoes one e-folding of growth. We can say that despite this instability, the jet is quite stable if it can propagate several (hundreds of) jet radii before the instability undergoes one e-folding of growth. We standardize these considerations by referring to jets with top-hat velocity profiles. If $r_j$ is the jet radius and $v_{z;\max}$ is the jet velocity, then the time taken by the jet to propagate $\chi$ jet radii is given by $T = \chi r_j / v_{z;\max}$. For a specified $\chi$, say $\chi = 400$, we say that the instability will not destabilize the jet if $\tau \geq T$. This is equivalent to saying that

$$\frac{\omega_I r_j}{c_s} \leq \frac{M}{\chi}$$

where "$M$" is the Mach number of the jet and $c_s$ is the sound speed in the jet. For this paper we consider jets with a Mach number of 4 and we use $\chi = 400$. I.e., a jet is said to be "quite stable" with respect to perturbations if the jet can propagate 400 jet radii before that perturbation undergoes growth by one e-folding. Comparison with the propagation of terrestrial jets, which destabilize in less than a 100 jet radii, indicates that our criterion is well-designed. Based on this criterion, a jet is quite stable if $\omega_I r_j / c_s \leq 10^{-2}$. For very highly magnetized jets, it is possible that the sound speed $c_s$ should be replaced by the Alfven speed of the jet material. However, in this work we don't consider jets that are very highly magnetized; besides, observers' biases usually favor jets that are closer to equipartition. Consequently, it is optimal to use just the sound speed.

**III) The Stability of the Fundamental Mode of the Pinch Instability**

Fig 3 shows the angular frequency (solid line) and temporal growth rate (dashed line) as a function of the wave number "$k$" for the fundamental mode of the pinch instability. Three panels are shown in Fig. 3. Fig. 3a corresponds to unmagnetized jets ($\beta = \infty$) and shows four different values of shear, corresponding to *a*=0, 0.3, 0.6 and 0.9. Fig. 3b corresponds to jets with equipartition between the thermal and magnetic pressure ($\beta = 1$) and the same four values of "*a*", corresponding to increasing shear from *a*=0 to *a*=0.9. Fig. 3c corresponds to jets with on-axis magnetic pressure that is twice as strong as the on-axis gas pressure ($\beta = 0.5$) and again the same four values of "*a*". In all, Fig. 3 shows the linear stability of the fundamental mode to pinch perturbations for twelve different models for sheared jets with various levels of magnetization. For values of wavenumber "*k*" with $\omega_I r_j / c_s \leq 10^{-2}$ the jets were assessed to be quite stable. This threshold is shown via a black dotted line in Fig. 3. (The same dotted line is also used for the remaining figures in this paper whenever we show a dispersion analysis.) Using our threshold of $\omega_I r_j / c_s \leq 10^{-2}$ it is very easy to see from Figs. 3a, 3b and 3c that increasing shear makes the jets quite stable to fundamental modes of the pinch instability for a substantial range of short and long wave lengths.

Fig. 3a, corresponding to four unmagnetized jets with increasing shear, makes the previous point very vividly. We see that with increasing shear the short wavelength modes as well as long wavelength modes are increasingly stabilized. Notice from Fig. 3a that all the fundamental pinch modes are unstable at $kr_j = 0.6$. We show this wave number by the short vertical arrow in Fig. 3a. This corresponds to a situation where the wavelength is about 10.5 jet radii. It has been anticipated by (Ferrari, Massaglia and Trussoni 1982) that short wavelengths are indeed stabilized by shear. The finding that long wavelengths are also stabilized by shear is indeed novel and based on our detailed stability analysis.

Payne and Cohn (1985) have already used a visualization of the pressure to develop further insight into jet stability. They were able to show that the pressure fluctuations give us new insights into the nature of fundamental and reflection modes. Since that early paper, few authors seem to have used the pressure variable to gain insights on jet stability. We have found that the fluctuations in the pressure variable give us substantial further insight into the stabilizing role of shear and certain configurations of magnetic fields. Our presentation is novel in the sense that we visualize this variable for jets with different levels of shear in a way where the variables can be directly intercompared. This allows us to obtain a deeper understanding of the role of increasing shear in stabilizing the jets for a given mode of oscillation. In this section, we visualize the fluctuations in the pressure variable for fundamental modes of the pinch instability. We do this in subsequent sections for each of the different modes of oscillation of the jet that are of interest to us.

To obtain a physical understanding of the results from Fig. 3, let us look at Fig. 4. Fig. 4 shows the pressure fluctuation in the jet and its ambient when the boundary of the jet has a 20% radial fluctuation. We show iso-pressure perturbation contours in Fig. 4 where the values of the contour lines can be obtained by matching the color of the contour to the color bar to the right of each figure. Since the undulations in the jets' boundary are communicated to the ambient medium via sound waves, it is appropriate to look at the pressure variable in the two media (i.e., the jet and its ambient medium). The whole process of a jet undergoing dynamical instability can be thought of as a process of converting the beam energy of the jet into pressure and velocity fluctuations in the ambient medium. In that sense too, the pressure is the appropriate variable to image in order to arrive at a physical picture of jet stability. We show the maximally unstable modes with $kr_j = 0.6$. The perturbed boundary of the jet is also shown in Fig. 4, just to provide the reader with a point of physical reference. However, please note that the boundary perturbation is not set by linear stability analysis. We conjecture that if the boundary perturbation is set to a given value then the pressure perturbations in different jets with different values of shear "*a*" but with all other parameters held fixed can indeed be inter-compared. Consequently, Fig. 4a shows the pressure for an unmagnetized jet with *a*=0; Fig. 4b shows the pressure for an unmagnetized jet with *a*=0.3; Fig. 4c shows the pressure for an unmagnetized jet with *a*=0.6; lastly, Fig. 4d shows the pressure for an unmagnetized jet with *a*=0.9. The Mach number, density ratio and perturbed wavelength are indeed the same for all four panels in Fig. 4. The color scale is indeed the same for all the panels in Fig. 4 so that the pressure contours can be inter-compared across all the panels within that figure. In other words, notice that the range of values associated with the four color bars is the same for the four panels in Fig. 4.

Fig. 4a shows that the pressure perturbations are strong within a jet with a top-hat velocity profile. Furthermore, they are organized like sound waves in a tube. Please focus on the jet's channel in this paragraph; i.e. please focus on the fluid inside the jet. The inclusion of even a mild shear in Fig. 4b changes the picture. We now see that the pressure perturbations are concentrated at the center of the jet. However, the boundary of the jet's channel experiences smaller perturbations. Fig. 4c corresponds to a jet with a modest shear and we see that the pressure perturbations are strongest in a smaller region inside the jet's channel. Fig. 4d corresponds to a jet with very strong shear and we see that the strongest pressure perturbations in that jet are confined to a narrow region along the jet's axis. Since the outer regions of the jet's channel are strongly sheared, we see that the pressure fluctuations are washed out by the presence of the shear. Realize from Fig. 2 that even the strongly sheared jet has a central velocity profile that does not experience much shear. Consequently, it is almost as if the flow close to the jet's axis in Fig. 4d forms a core of a jet with a higher and flatter velocity and the pressure fluctuations within the jet's channel are concentrated on-axis. We also invite the reader to cross compare the colors for the pressure inside the jets' channel in Figs. 4c and 4d to the corresponding colors for the pressure inside the jets' channel in Figs. 4a and 4b. By comparing the colors, as well as the range of colors, in the different panels in Fig. 4, we see that the pressure fluctuations inside the jets' channel are larger in the unsheared and mildly sheared configurations than in the strongly sheared configurations.

Eventually, a stability analysis of jets should lead us to a better understanding of how the jet loses energy to its ambient medium and, therefore, destabilizes. In the previous paragraph we focused exclusively on the pressure fluctuations within the jets' channel. Let us now revisit Fig. 4 with an emphasis on the pressure fluctuations outside the jets' channel, i.e. in the ambient medium. The previous paragraph has shown us that increasing shear results in smaller pressure fluctuations reaching the boundary of the jet. These pressure fluctuations should trigger pressure perturbations in the ambient medium. We now see that the pressure fluctuations in the ambient medium are stronger for the unsheared and mildly sheared jets in Figs. 4a and 4b respectively. We also see that the pressure fluctuations in the ambient medium are weaker for the modest and strongly sheared jets in Figs. 4c and 4d. Therefore, we understand that the fundamental pinch modes of jets that are strongly sheared lose less sound energy to their ambient medium than jets that have mild or no shear. Clearly, this is a far-reaching insight which enables us to understand why long wavelength and also short wavelength modes were stabilized in Fig. 3a with increasing shear. Observe too from Fig. 3b that the jets with equipartition magnetic field also show all the same trends as Fig. 3a. Comparing Fig. 3b to Fig. 3a, which Fig. 4 in the backdrop, has made it easy to understand the trends in Fig. 3b and we do not show the corresponding pressure profiles.

Fig. 3c shows the fundamental modes for the sheared and strongly magnetized jets. By comparing the *a*=0.3, 0.6 and 0.9 cases we see an overall trend towards increasing stabilization with increasing shear. By comparing these cases with the a=0 case in Fig. 3c, we see that the strong magnetic field is effective in stabilizing the short wavelength modes. This makes sense because the magnetic field that we used has a strong toroidal component. However, the magnetic field is not effective in stabilizing the long wavelength modes, which are usually more damaging to jet stability. Notice from Fig. 3c that all the fundamental pinch modes are unstable at $kr_j = 0.3$. We show this wave number by the short vertical arrow in Fig. 3c. In the next paragraph we analyze the pressure fluctuations for the four jets from Fig. 3c at $kr_j = 0.3$.

Fig. 5 shows the pressure perturbations in the four jets from Fig. 3c for an unstable wavenumber given by $kr_j = 0.3$. I.e., please realize that this is a longer wavelength than the one shown in Fig. 4. Fig. 5a corresponds to a top-hat velocity profile. Even so, we see that the pressure fluctuations within the jet's channel are confined to a central region. Please compare Fig. 5a for the magnetized jet to Fig. 4a for the unmagnetized jet and notice that the pressure fluctuations in Fig. 5a are confined to a small region close to the jet's axis. To understand why, please focus on the magnetic field structure in Fig. 1. We see that the toroidal component of the magnetic field becomes stronger at larger jet radii. This confines the pressure fluctuations to a smaller fraction of the jet's channel. Fig. 5b shows the same trends as Fig. 5a. Fig. 5c and 5d correspond to jets with increasing shear; they show even smaller pressure fluctuations. We see now that the combination of strong magnetic field and strong shear has resulted in milder pressure fluctuations both within the jets' channel as well as in the ambient region. We therefore see that the combination of increasing magnetic field and increasing shear has a strongly stabilizing influence on the fundamental pinch modes of the light, high Mach number jets studied here.

**IV) The Stability of the First Reflection Mode of the Pinch Instability**

Fig 6 shows the angular frequency (solid line) and temporal growth rate (dashed line) as a function of the wave number "*k*" for the first reflection mode of the pinch instability. Three panels are shown in Fig. 6. Fig. 6a corresponds to unmagnetized jets ($\beta = \infty$) and shows four different values of shear, corresponding to *a*=0, 0.3, 0.6 and 0.9. Fig. 6b corresponds to jets with equipartition between the thermal and magnetic pressure ($\beta = 1$) and the same four values of "*a*", corresponding to increasing shear from *a*=0 to *a*=0.9. Fig. 6c corresponds to jets with on-axis magnetic pressure that is twice as strong as the on-axis gas pressure ($\beta = 0.5$) and again the same four values of "*a*". In all, Fig. 6 shows the linear stability of the first reflection mode to pinch perturbations for twelve different models for sheared jets with various levels of magnetization. Using our threshold of $\omega_I r_j / c_s \leq 10^{-2}$ it is very easy to see from Figs. 6a, 6b and 6c that increasing shear makes the jets quite stable to the first reflection modes of the pinch instability for a substantial range of short wave lengths. The nature of the first reflection mode is such that it does not destabilize the jet at long wave lengths anyway. So, when considering reflection modes, we are only interested in enhanced stabilization of short wavelength modes.

Realize, at the onset, that the reflection modes only destabilize the jets at shorter wavelengths. Consequently, we can only talk about improved stability at short wavelengths when considering reflection modes. Please cross-compare Figs. 3a and 6a. We see that increasing shear results in a smaller island of instability for both the fundamental mode and the first reflection mode of the pinch instability. Cross-comparing Figs. 3b and 6b shows the same trend continues when the magnetic pressure is in equipartition with the gas pressure. In Fig. 6c we see most of that trend continued. However, we see, as before, that the presence of shear is more effective at stabilizing short wavelength modes than the presence of a strong magnetic field.

The pressure fluctuations in the first reflection modes of the pinch instability are also shown for the unmagnetized jets with wave number $kr_j = 1$ in Fig. 7. As in Fig. 4, we see that increasing shear causes the pressure fluctuations within the jets' channel to be concentrated closer to the axis of the jet. I.e., observe that the contour intervals have a smaller spacing in the radial direction in-close to the jets' axis and that this trend increases with increasing shear. In Fig. 7d we see that the outer parts of the jet's channel are mostly free of pressure fluctuations. Only the part of the jet that is closest to the jet's axis has a concentration of pressure contours indicating that all the pressure fluctuations in Fig. 7d are concentrated close to the jet's axis. We can also turn our attention to the pressure fluctuations in the ambient medium that surrounds the jets in Fig. 7. The strongly sheared jet in Fig. 7d shows that the sound wave fluctuations that propagate into the ambient medium are weaker than the sound waves in the ambient media of Figs. 7a, 7b and 7c. We, therefore, conclude that the presence of shear reduces the amount of energy that is imparted to the ambient medium by the first reflection modes of the pinch instability.

The pressure fluctuations in the first reflection modes of the pinch instability are also shown for the strongly magnetized jets with wave number $kr_j = 1.5$ in Fig. 8. Please compare Fig. 8 to Fig. 7 to realize that all the trends that we saw in the unmagnetized jets are also present in the strongly magnetized jets.

Our overall conclusion from Sections III and IV is that shear has a stabilizing influence on both the long and the short wavelength modes for the pinch instability. This enhanced stability extends to fundamental as well as first reflection modes. It may, therefore, be expected that the second and higher reflection modes of the pinch instability show similar stabilization. We have also shown that for the same perturbation amplitude, the strongly sheared jets impart a smaller fraction of the jet's kinetic energy to the ambient medium.

**V) The Stability of the Fundamental Mode of the Kink Instability**

Fig 9 shows the angular frequency (solid line) and temporal growth rate (dashed line) as a function of the wave number "$k$" for the fundamental mode of the kink instability. Three panels are shown in Fig. 9. Fig. 9a corresponds to unmagnetized jets ($\beta = \infty$) and shows four different values of shear, corresponding to $a$=0, 0.3, 0.6 and 0.9. Fig. 9b corresponds to jets with equipartition between the thermal and magnetic pressure ($\beta = 1$) and the same four values of "$a$", corresponding to increasing shear from $a$=0 to $a$=0.9. Fig. 9c corresponds to jets with on-axis magnetic pressure that is twice as strong as the on-axis gas pressure ($\beta = 0.5$) and again the same four values of "$a$". In all, Fig. 9 shows the linear stability of the fundamental mode to kink perturbations for twelve different models for sheared jets with various levels of magnetization.

Using our threshold of $\omega_I r_j / c_s \leq 10^{-2}$ it is very easy to see from Figs. 9a, 9b and 9c that increasing shear makes the jets quite stable to fundamental modes of the kink instability for a substantial range of short wave lengths.

It is well-known in the literature (Begelman 1998, Lyubarskii 1999, Appl *et al.* 2000) that the long wavelength modes of the kink instability can be more destabilizing (destructive) to jets than the pinch instability. In Fig. 3 we showed that the long as well as the short wavelength modes of the pinch instability are progressively stabilized by increasing shear. From Fig. 9 we see that the long wavelength modes of the kink instability are less affected by the presence of shear. In all three panels of Fig. 9 we see that only the *a*=0.9 jet shows improved stability at long wavelengths because of the presence of shear. Even then, the improvement is only by a small factor. The long wavelength kink instability is a body mode, i.e. the entire body of the jet is displaced from one side of the axis to another. As a result, the internal shear cannot do much to stabilize it.

From Figs. 9a and 9b we see that the fundamental mode of the kink instability is indeed strongly stabilized at short wavelengths due to the presence of increasing shear. Fig. 9c shows that the strong magnetic field competes with the shear at short wavelength in stabilizing the fundamental mode of the kink instability. However, even in Fig. 9c we see that the strongly sheared jet shows improved stability at short wavelengths compared to its less sheared counterparts. By comparing Figs. 9a and 9c we can certainly conclude that a combination of strong shear and strong magnetic field certainly has a very stabilizing influence on short wavelength fluctuations of the fundamental mode of the kink instability.

Fig. 10 shows the pressure fluctuation in the jet and its ambient medium when the boundary of the jet has a 20% radial fluctuation. We show the maximally unstable modes with $kr_j = 0.6$. The perturbed boundary of the jet is also shown in Fig. 10, just to provide the reader with a point of physical reference. It is worthwhile to compare the pressure contours in Fig. 10. As with the fundamental pinch modes shown in Fig. 4, we see that increasing shear causes the pressure contours in the fundamental kink modes to become increasingly concentrated towards the axis of the jet. Furthermore, by noting the similarity between Fig. 4d and Fig. 10d we see that the strongly sheared jet produces smaller pressure fluctuations in the ambient medium. As a result, a smaller fraction of the jets' beam energy is conveyed to the ambient medium as the jets' axial velocity profile becomes increasingly sheared.

Fig. 11 shows the role of increasing shear in the stability of a strongly magnetized jet. As with Fig. 5, we see that the presence of a strong magnetic field concentrates the pressure perturbations in the fundamental mode of the kink instability increasingly towards the high-velocity core of the jet; i.e., towards its axis. Comparing Fig. 11d to Fig. 11a we can again see that the strongly sheared jet produces smaller pressure fluctuations in the ambient medium.

**VI) The Stability of the First Reflection Mode of the Kink Instability**

Fig 12 shows the angular frequency (solid line) and temporal growth rate (dashed line) as a function of the wave number "*k*" for the first reflection mode of the kink instability. Three panels are shown in Fig. 12. Fig. 12a corresponds to unmagnetized jets ( $\beta = \infty$ ) and shows four different values of shear, corresponding to *a*=0, 0.3, 0.6 and 0.9. Fig. 12b corresponds to jets with equipartition between the thermal and magnetic pressure ($\beta = 1$) and the same four values of "*a*", corresponding to increasing shear from *a*=0 to *a*=0.9. Fig. 12c corresponds to jets with on-axis magnetic pressure that is twice as strong as the on-axis gas pressure ($\beta = 0.5$) and again the same four values of "*a*". In all, Fig. 12 shows the linear stability of the first reflection mode to kink perturbations for twelve different models for sheared jets with various levels of magnetization. Using our threshold of $\omega_I r_j / c_s \leq 10^{-2}$ it is very easy to see from Figs. 12a, 12b and 12c that increasing shear makes the jets quite stable to first reflection modes of the kink instability for a substantial range of short wave lengths.

Figs. 13 and 14 show the pressure fluctuations in the first reflection mode of the kink instability from the panels in Fig. 12a and 12c. As in Section IV, we see that increasing shear only helps in stabilizing the reflection modes of the jets at shorter wavelengths. To see that, please compare Fig. 13d to Fig. 13a. Alternatively, please compare Fig. 14d to Fig. 14a.

Our overall conclusion from Sections V and VI is that shear has a stabilizing influence for the short wavelength modes of the kink instability. The fundamental mode of the kink instability is not stabilized too much on the longest wavelengths. (This stands in contrast to the fundamental mode of the pinch instability which was, indeed, stabilized quite substantially even at long wavelengths.) This enhanced stability at short wavelengths extends to fundamental as well as reflection modes of the kink instability. We have also shown that for the same perturbation amplitude, the strongly sheared jets impart lower amount of their beam energy to the ambient medium.

**VII) Discussion and Conclusions**

Many types of astrophysical jets, e.g. AGN jets and jets from young stars, demonstrate the remarkable ability to survive over length scales spanning many orders of magnitude. This is in stark contrast with the terrestrial jets which become destroys by instabilities over few tens to hundreds of their radius. Numerous analytical and numerical studies of jet stability have been conducted in attempts to find the explanation of the observation but no widely-accepted conclusion of this issue has been reached yet.

Most previous analytic studies of linear stability were dealing with oversimplified jet structure, which was demanded by the need to make the mathematical problem treatable. More realistic configurations can be analyzed only with the help of numerical approach. In Kim et al. (2015), we described a robust numerical method for studying jets stability, and used it to study non-relativistic magnetized jets with realistic magnetic field structure. In the present paper, we expanded this study by considering the role of the velocity share. The pinch and kink modes, both fundamental and first reflection modes, have been considered.

Overall, we find that velocity shear plays a stabilizing role, by narrowing the range unstable modes and reducing their growth rates. The effect is particularly strong for the pinch modes, which become suppressed both at short and long wavelengths, with only a narrow unstable range remaining for strong shear. However, for the kink modes the effect is weaker and only short (in comparison with the jet radius) wavelength modes, are suppressed. The long-wavelength modes, which are most threatening the jet disintegration, are not influenced by the share. Because such long-wavelength spectral components of perturbations are expected to be present in astrophysical jets due to a variety of reasons, this conclusion is a matter of concern. According to our results, the amplitude of the fastest growing kink modes e-folds on the length scale $l_e \sim 3Mr_j$, where $M$ is the jet Mach number.

Our results, as well as those presented in Kim et al. (2015), suggest that although the details of the inner structure of jets make an impact on the jet stability, taken alone they cannot explain the observations of astrophysical jets, which can propagate over the distances exceeding their initial radius by more than a million times. Other important factors have to play a part. One possibility is hinted by the observed rapid lateral expansion of astrophysical jets, which appear rather more conical or parabolic than cylindrical. The lateral expansion tends to slow down the growth of unstable modes simply because it increases the communication time across the jet (e.g. Rosen & Hardee 2000; Moll, Spruit & Obergaulinger 2008; Porth & Komissarov 2015). Given the rapidly declining external pressure in the surrounding of many astrophysical jet engines, the jets may even become freely expanding, which totally suppresses global instabilities. When jets enter flat section of the external pressure distribution they may re-confine and then develop instabilities. Porth & Komissarov (2015) argue that this is how FR-I jets turn into subsonic turbulent plumes on kpc-scales.

Instabilities may be required to turn on the emission of astrophysical jets, converting part of their bulk motion or large-scale magnetic energy into the kinetic energy of emitting particles. Some kind of dissipation and in-situ particle acceleration is required when the life-time of emitting particles is small compared to the jet travel time. This is indeed the case for the high-energy synchrotron electrons in many AGN jets. If they are energized via jet instabilities, these are likely to be local and hence non-threatening to the jet integrity. Porth & Komissarov (2015) have demonstrated that such local instabilities may develop in the jet core, which expands at much slower rate compared to the whole jet. They argued that this may result only in the central part of the jets shining brightly, whereas their extended sheaths remaining rather dim. Our results support this possibility. They show that in the presence of velocity shear the perturbations develop profiles (eigen-functions) strongly peaked towards the axis of the jet. Although the non-linear evolution may strongly deviate from the prediction of the linear theory, this finding suggests stronger dissipation and higher emissivity near the jet axis. Compared to FR-I jets, several FR-II jets do indeed exhibit a core-brightened structure (Bridle & Perley 1984, Bicknell 1984).

The present study deals only with the linear stability of non-relativistic jets as the relativistic equations are more complicated and substantially harder to solve. However, most AGN jets and GRB jets are relativistic and we are planning to extend our formulation to relativistic jets in the near future.

**Acknowledgements**

DSB acknowledges support via NSF grants NSF-ACI-1307369, NSF-DMS-1361197 and NSF-ACI-1533850. DSB and ML also acknowledge support via NASA grants from the Fermi program as well as NASA-NNX 12A088G. Several simulations were performed on a cluster at UND that is run by the Center for Research Computing. Computer support on NSF's XSEDE and Blue Waters computing resources is also acknowledged.

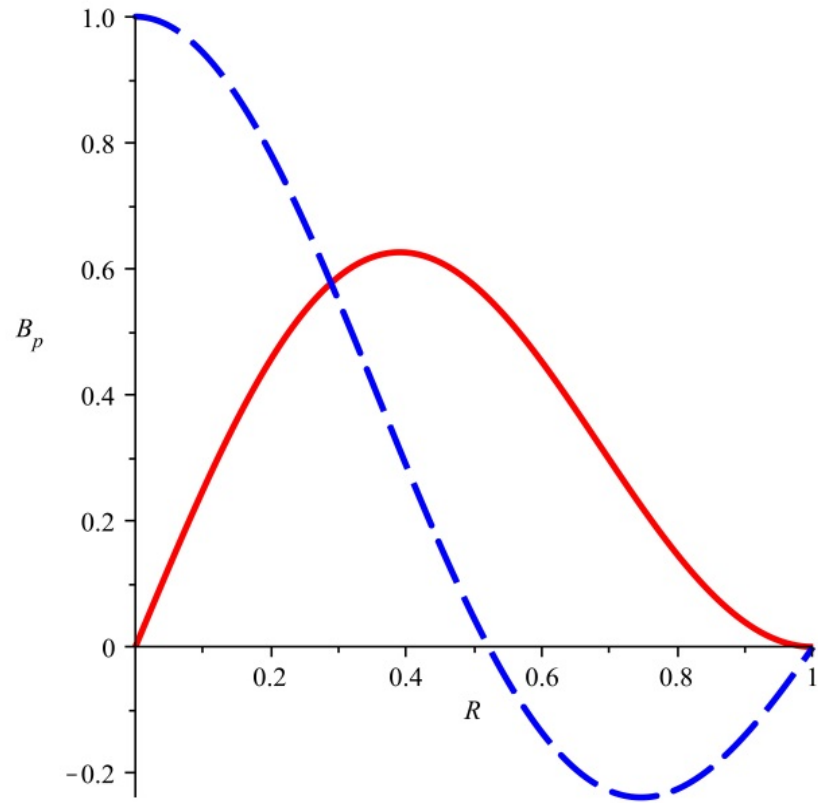

*Fig 1, from Gourgouliatos et al (2012), shows the toroidal magnetic field (red solid line) and the axial field (blue dashed line) as a function of the jet radius. Notice that the fields are zero at the jet boundary, resulting in jets that do not have a current sheet at the boundary.*

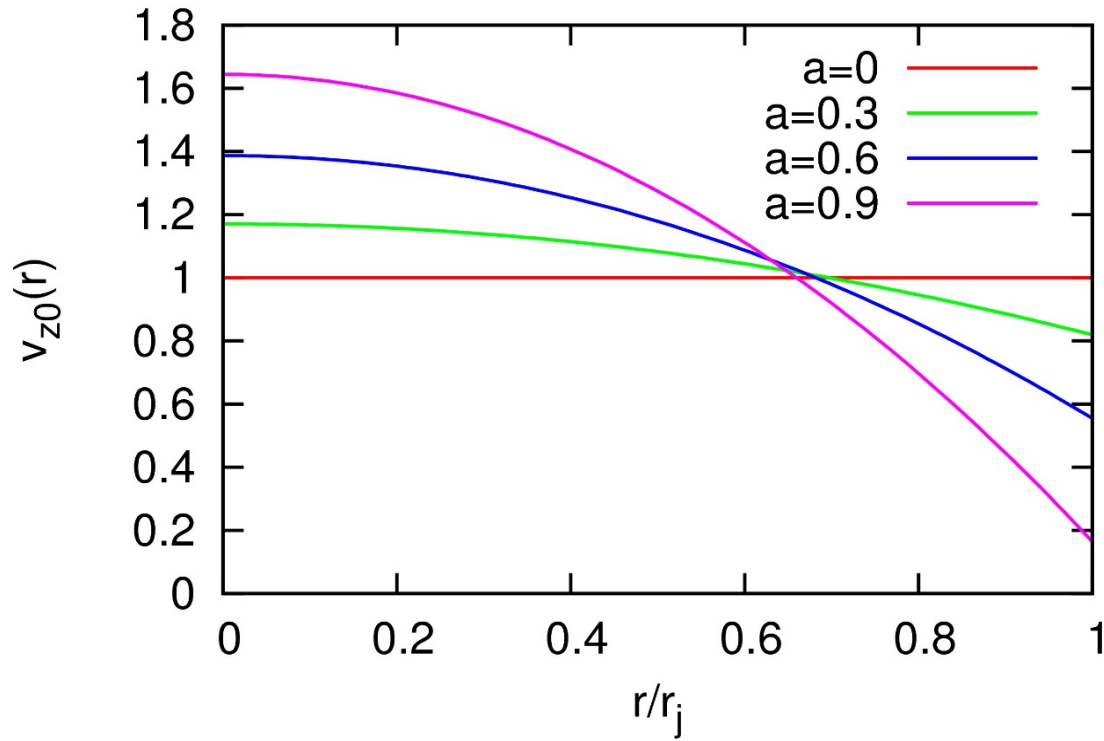

Fig. 2 shows velocity profile of the jet as a function of scaled radius. Notice that even for the hydrodynamic jet, the Mach number is not a constant for non-zero values of "a". For the magnetized jet, the jet density and pressure can also very as a function of radius. As a result, the connection with a single "Mach number" becomes even more tenuous. For want of an alternative, we still parametrize jet properties w.r.t. the on-axis Mach number.

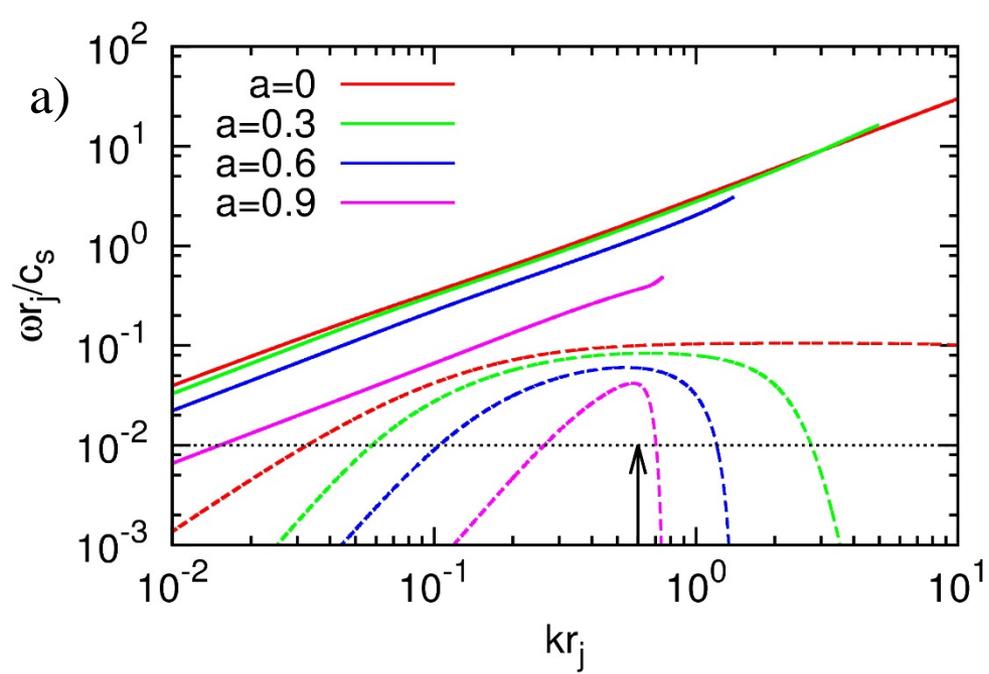
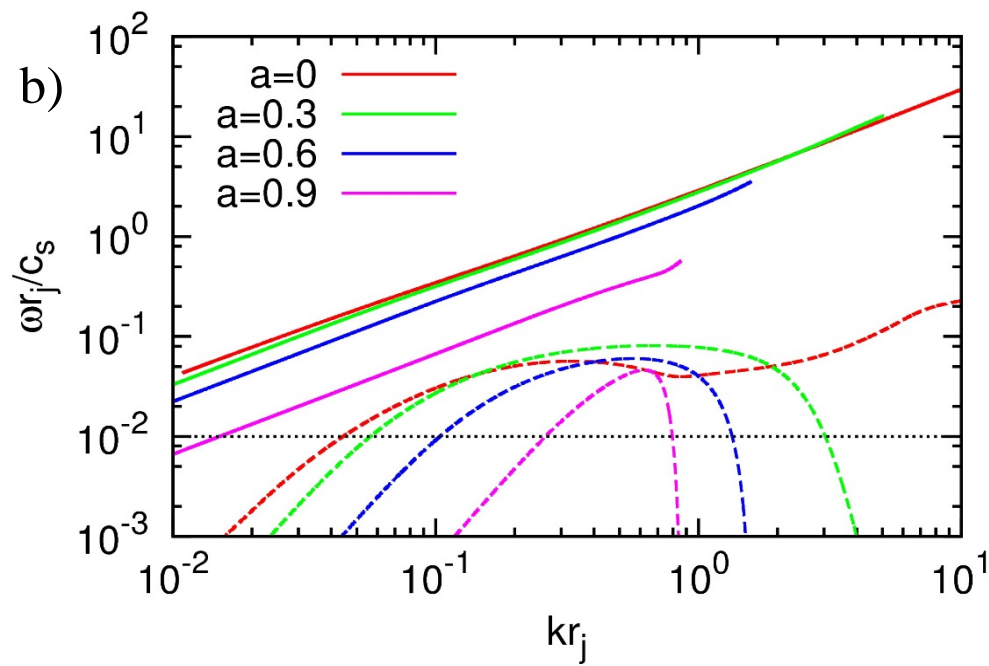

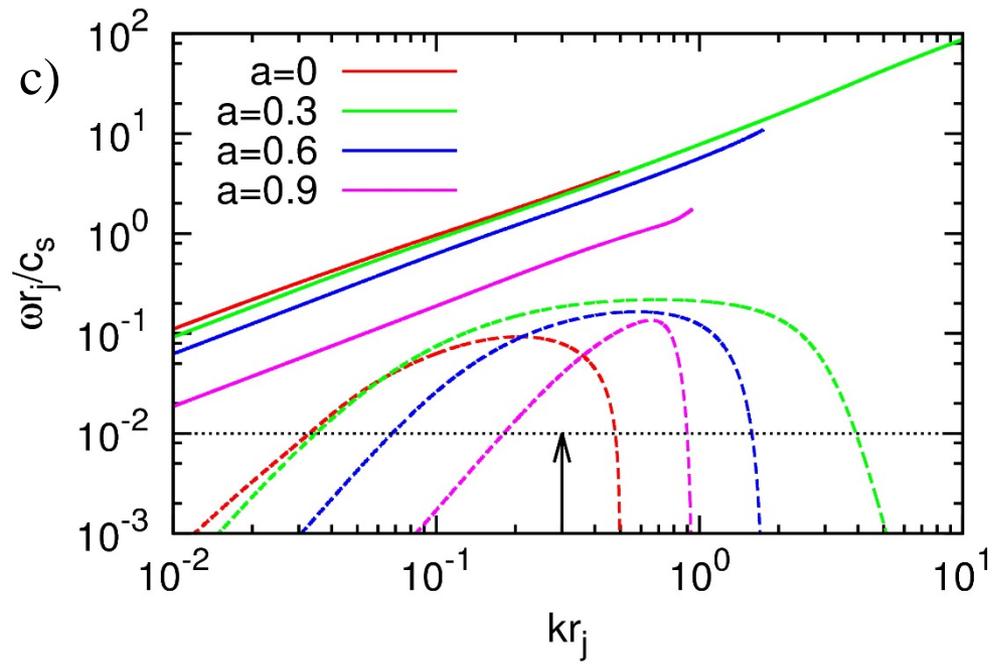

Fig. 3 shows the angular frequency (solid line) and temporal growth rate (dashed line) versus longitudinal wavenumber k for pinching (m=0) fundamental mode of a non-magnetized jet. In Fig. 3a, the jet has M=4 and η=0.1. Increasing values of the parameter "a" indicate increasing shear, with a=0 (no shear) to a=0.9 (maximal shear). Fig. 3b shows the same information for a jet with β=1 (i.e. magnetic field is in equipartition with gas pressure). Fig. 3c shows the same information for a jet with β=0.5 (magnetically dominated).

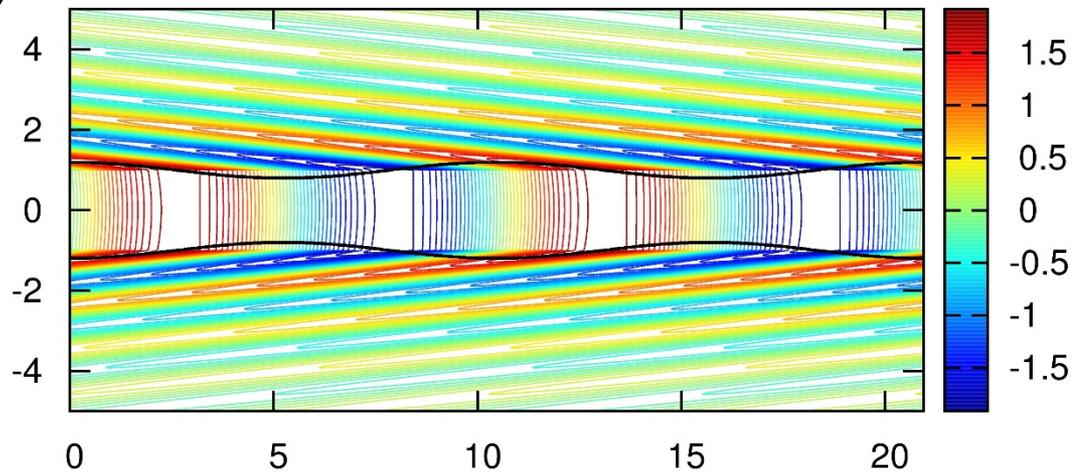

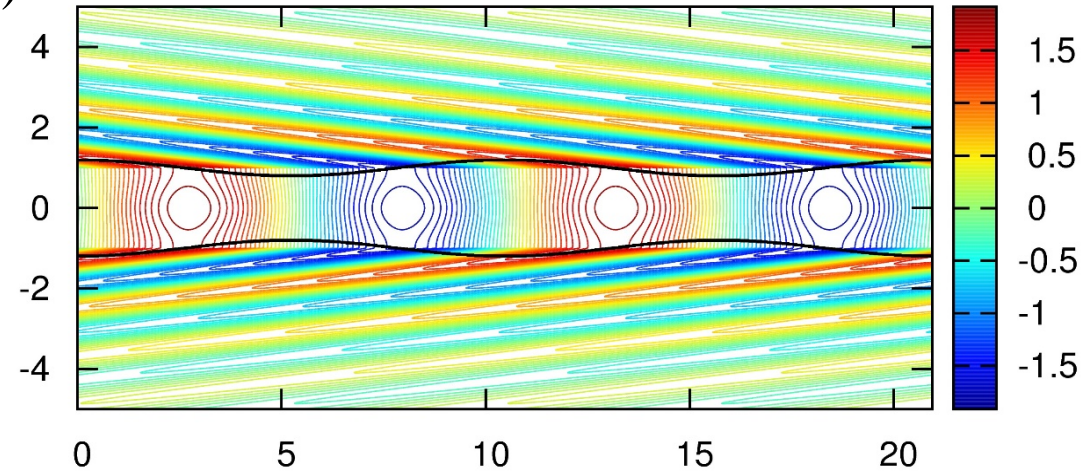

c) 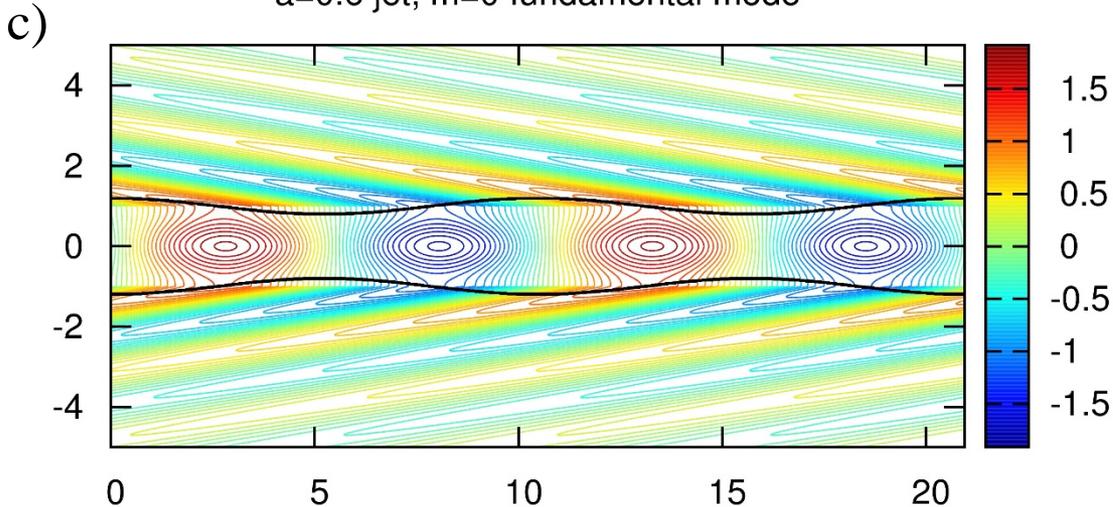

d) 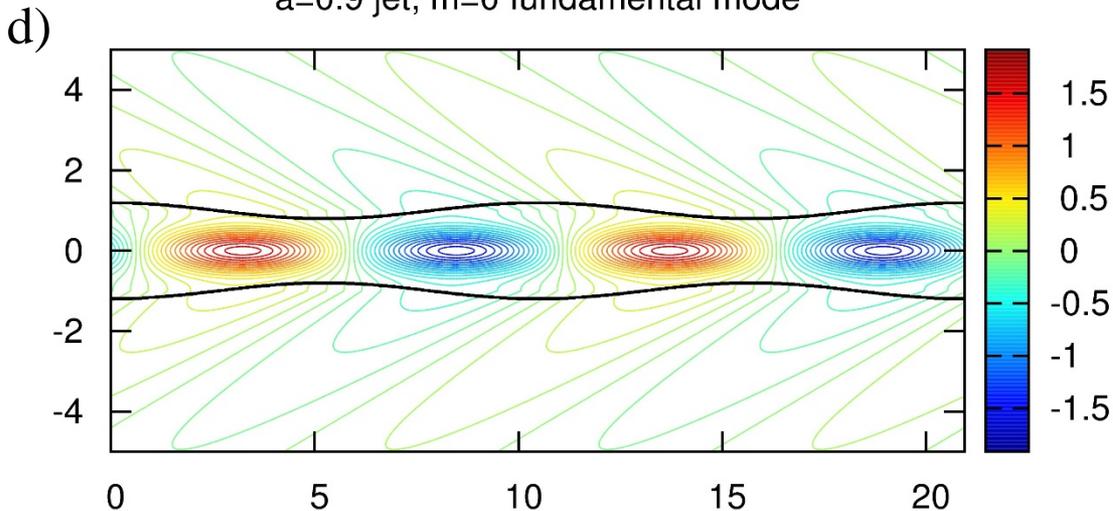

Fig. 4a shows the pressure variation in a non-magnetized jet with a top hat profile. Fig. 4b shows the pressure variation in a non-magnetized jet with a=0.3 (mild shear). Fig. 4c shows the pressure variation in a non-magnetized jet with a =0.6 (modest shear). Fig. 4d shows the pressure variation in a non-magnetized jet with a=0.9 (strong shear). For all the cases in Fig. 4 we have k $r_j$ = 0.6. In all the cases, the jet's boundary has a fluctuation that is 20% of the jet's radius. The pressures are all on the same scale so that the pressures across panels within a figure can be inter-compared.

a)

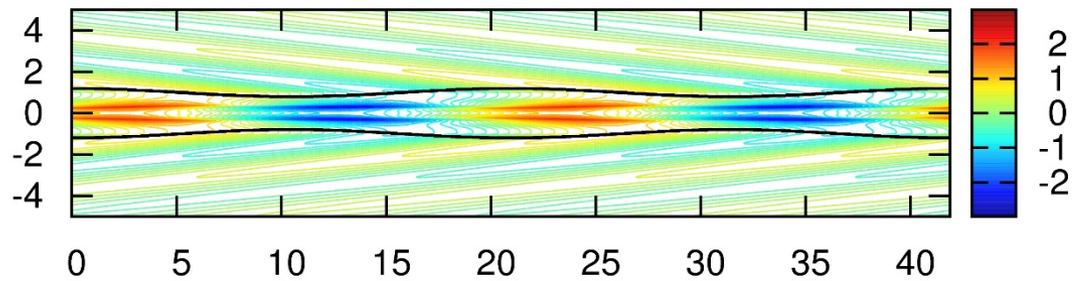

b)

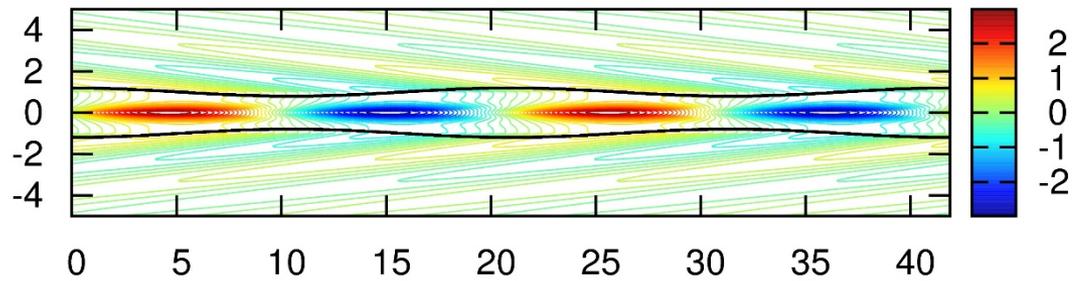

c)

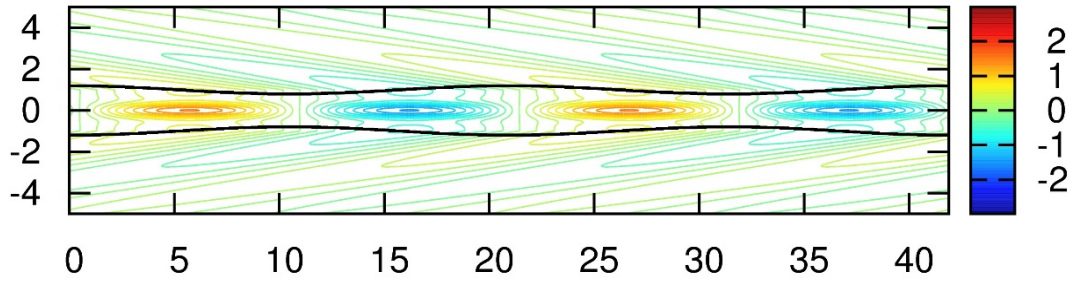

d)

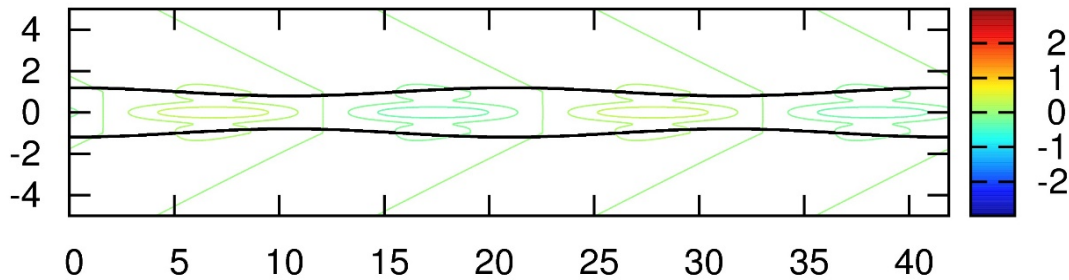

*Fig. 5a shows the pressure variation in a strongly-magnetized jet ($\beta=0.5$) with a top hat profile. Fig. 5b shows the pressure variation in a strongly-magnetized jet with $a=0.3$ (mild shear). Fig. 5c shows the pressure variation in a strongly-magnetized jet with $a=0.6$ (modest shear). Fig. 5d shows the pressure variation in a strongly-magnetized jet with $a=0.9$ (strong shear). For all the cases in Fig. 5 we have $k\,r_j = 0.3$. In all the cases, the jet's boundary has a fluctuation that is 20% of the jet's radius. The pressures are all on the same scale so that the pressures across panels within a figure can be inter-compared.*

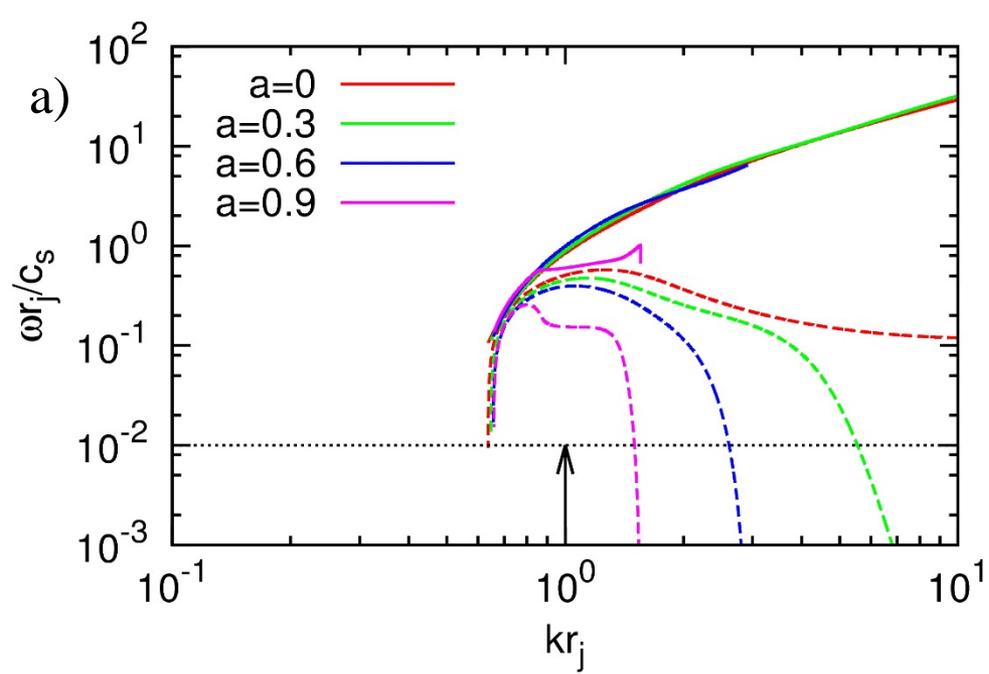
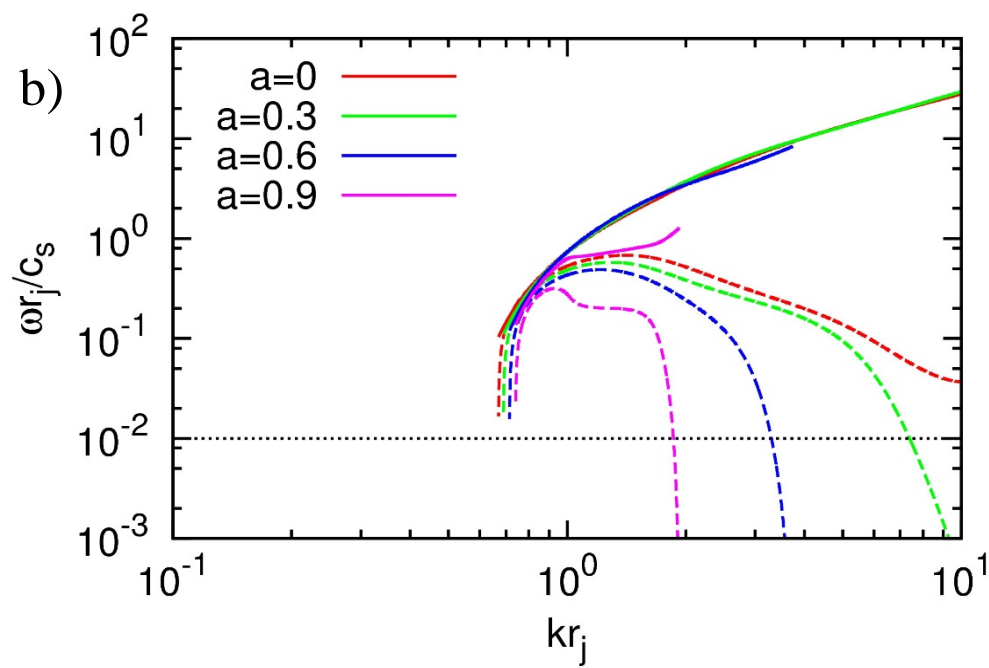

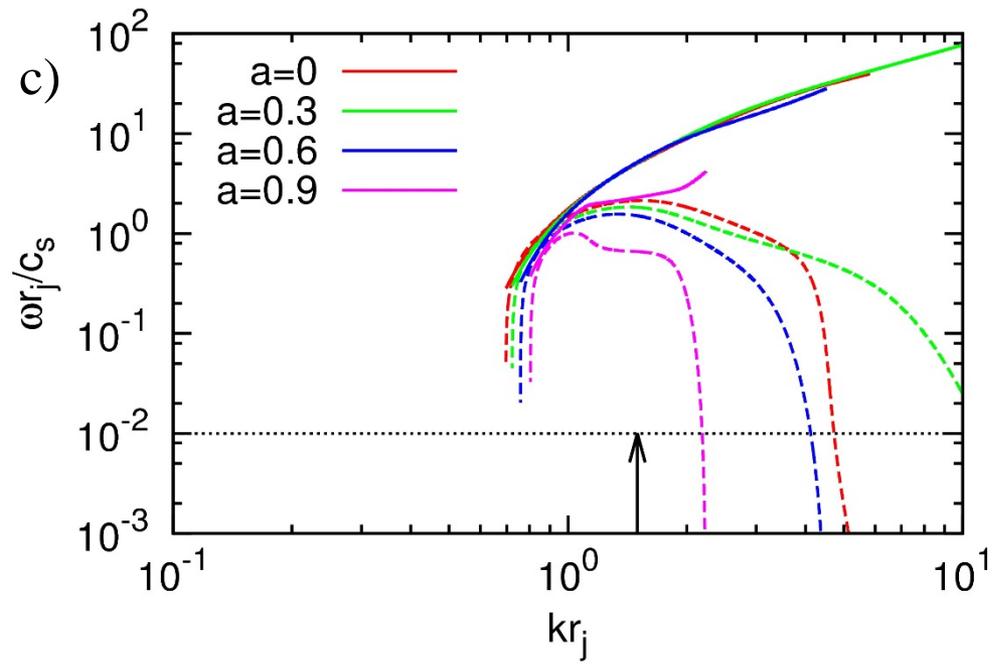

Fig. 6 shows the angular frequency (solid line) and temporal growth rate (dashed line) versus longitudinal wavenumber k for pinching (m=0) 1st reflection mode of a non-magnetized jet. In Fig. 6a, the jet has M=4 and η=0.1. Increasing values of the parameter "a" indicate increasing shear, with a=0 (no shear) to a=0.9 (maximal shear). Fig. 6b shows the same information for a jet with β=1 (i.e. magnetic field is in equipartition with gas pressure). Fig. 6c shows the same information for a jet with β=0.5 (magnetically dominated).

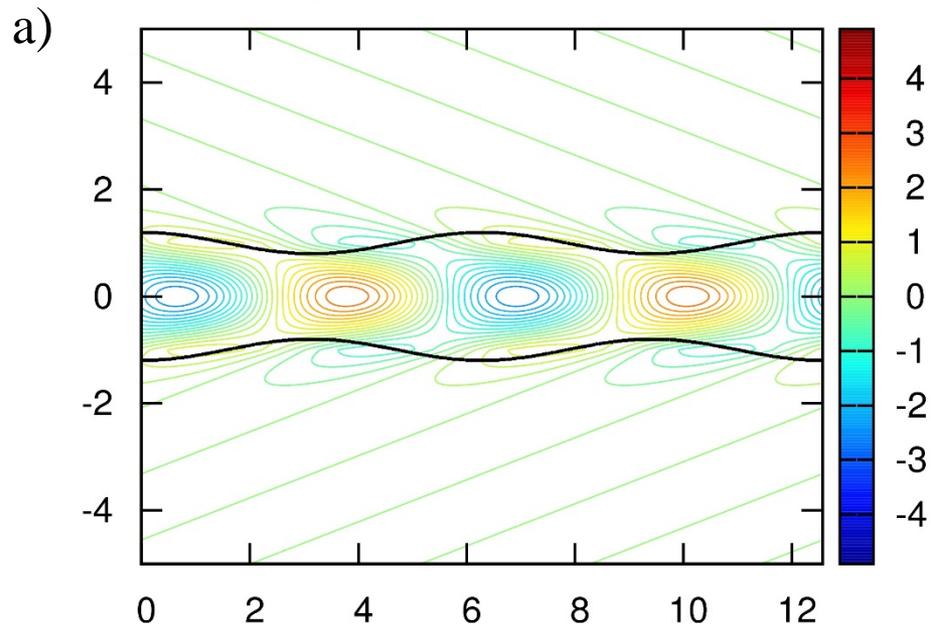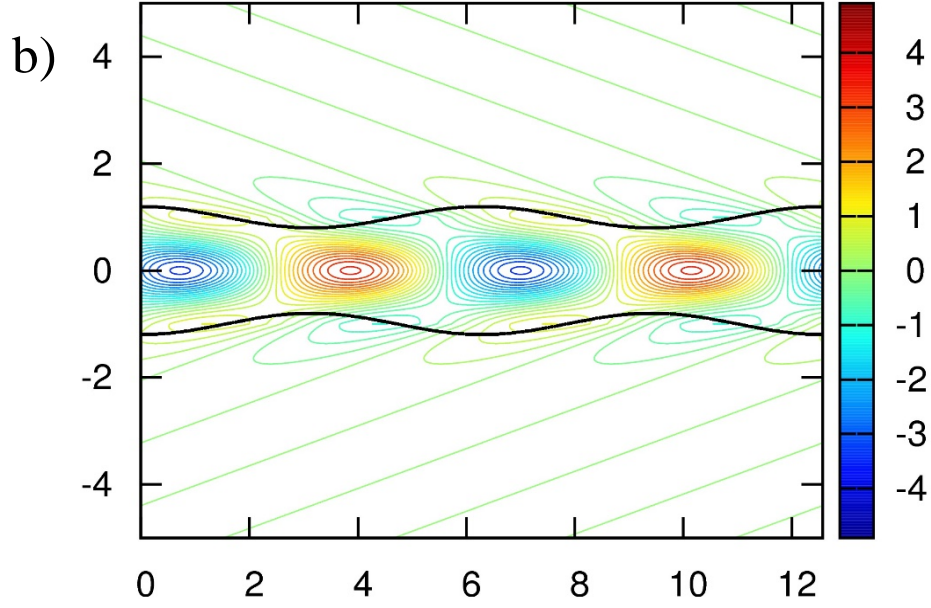

c) 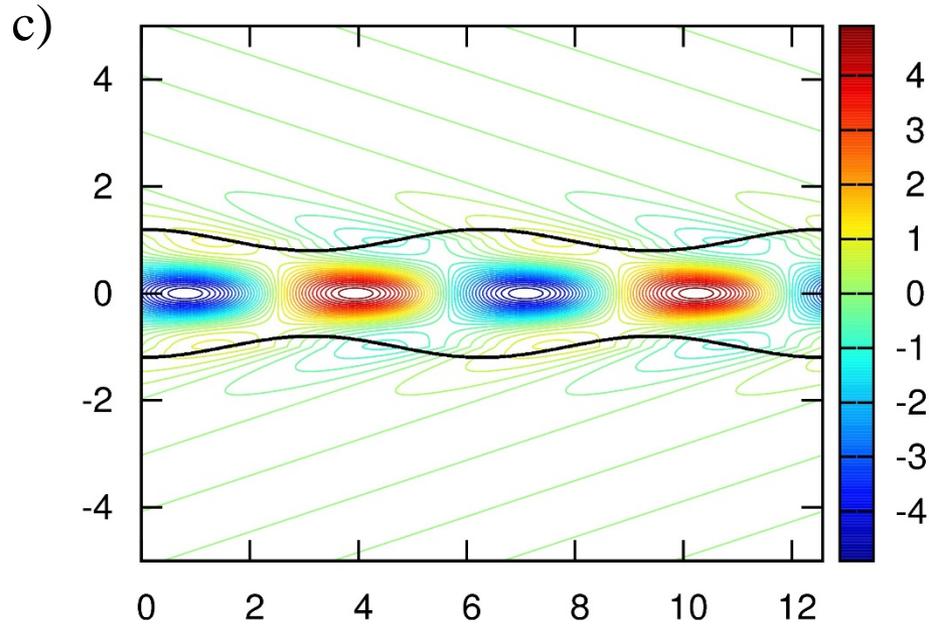

d) 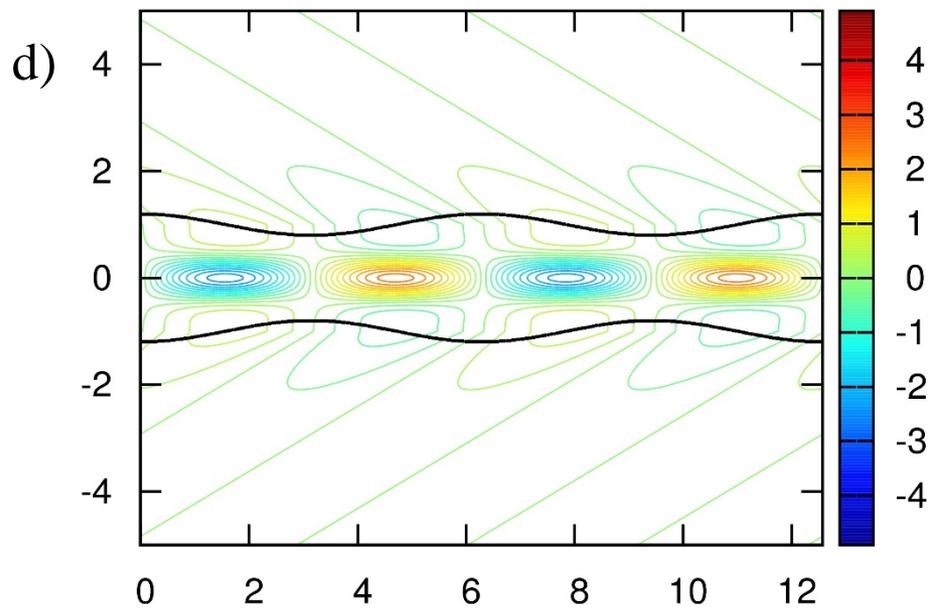

*Fig. 7a shows the pressure variation in a non-magnetized jet with a top hat profile. Fig. 7b shows the pressure variation in a non-magnetized jet with a=0.3 (mild shear). Fig. 7c shows the pressure variation in a non-magnetized jet with a =0.6 (modest shear). Fig. 7d shows the pressure variation in a non-magnetized jet with a=0.9 (strong shear). For all the cases in Fig. 7 we have k $r_j$ = 1. In all the cases, the jet's boundary has a fluctuation that is 20% of the jet's radius. The pressures are all on the same scale so that the pressures across panels within a figure can be inter-compared.*

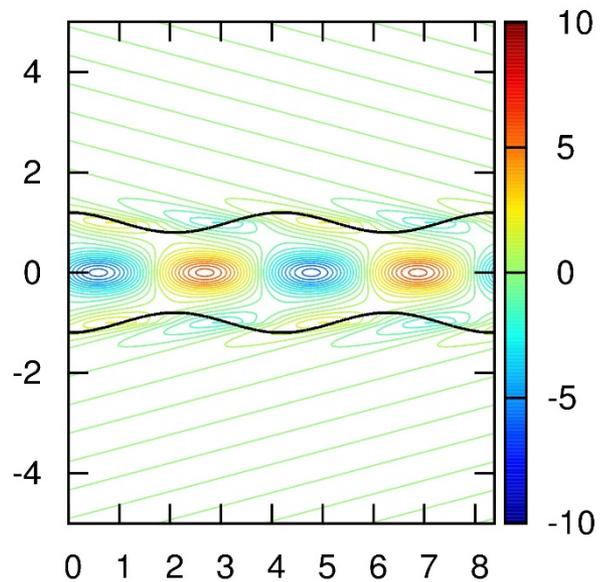

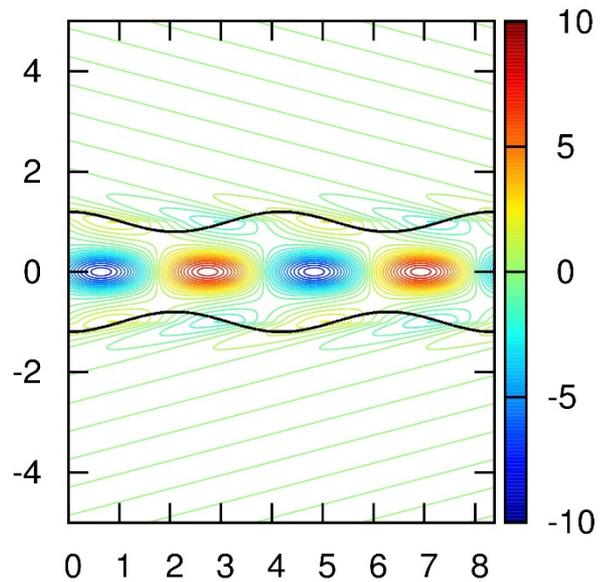

c) 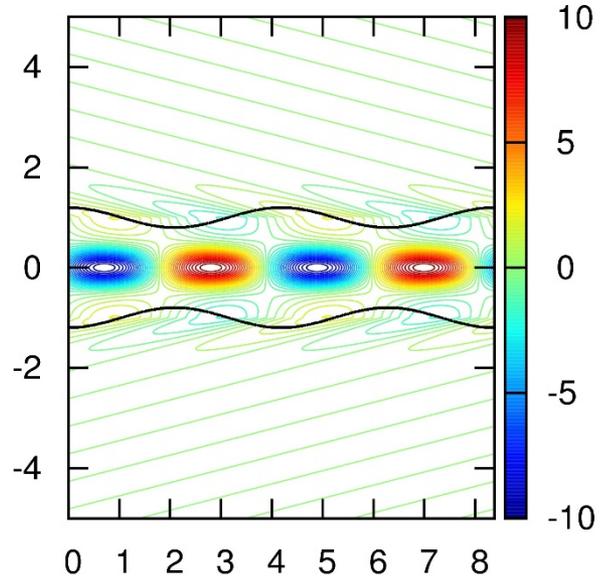

d) 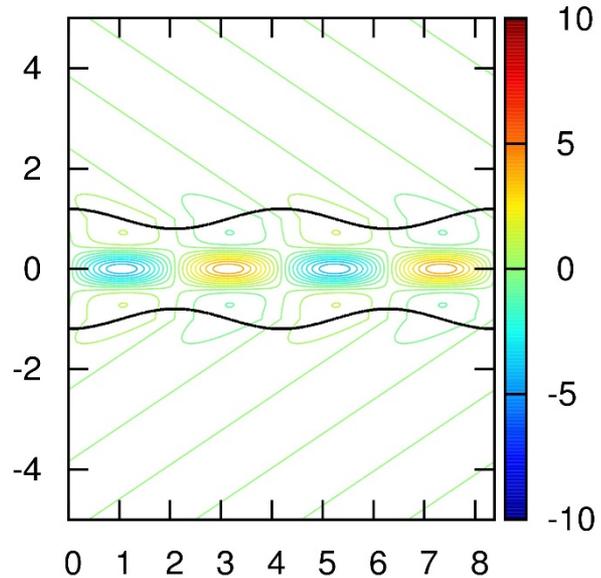

*Fig. 8a shows the pressure variation in a strongly-magnetized jet ($\beta=0.5$) with a top hat profile. Fig. 8b shows the pressure variation in a strongly-magnetized jet with a=0.3 (mild shear). Fig. 8c shows the pressure variation in a strongly-magnetized jet with a =0.6 (modest shear). Fig. 8d shows the pressure variation in a strongly-magnetized jet with a=0.9 (strong shear). For all the cases in Fig. 8 we have $k\,r_j = 1.5$. In all the cases, the jet's boundary has a fluctuation that is 20% of the jet's radius. The pressures are all on the same scale so that the pressures across panels within a figure can be inter-compared.*

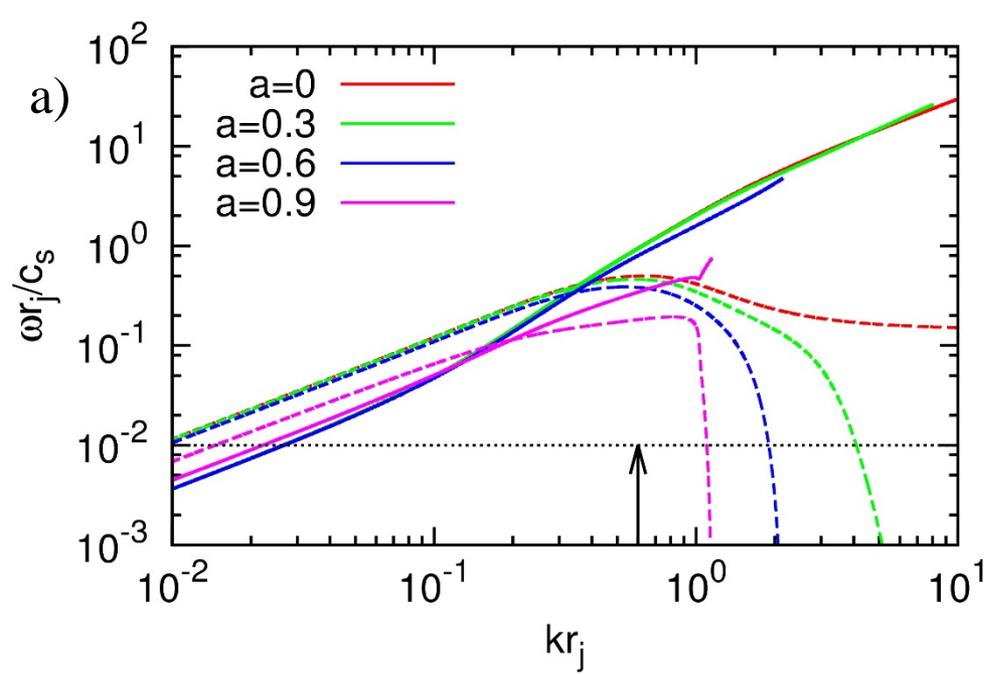

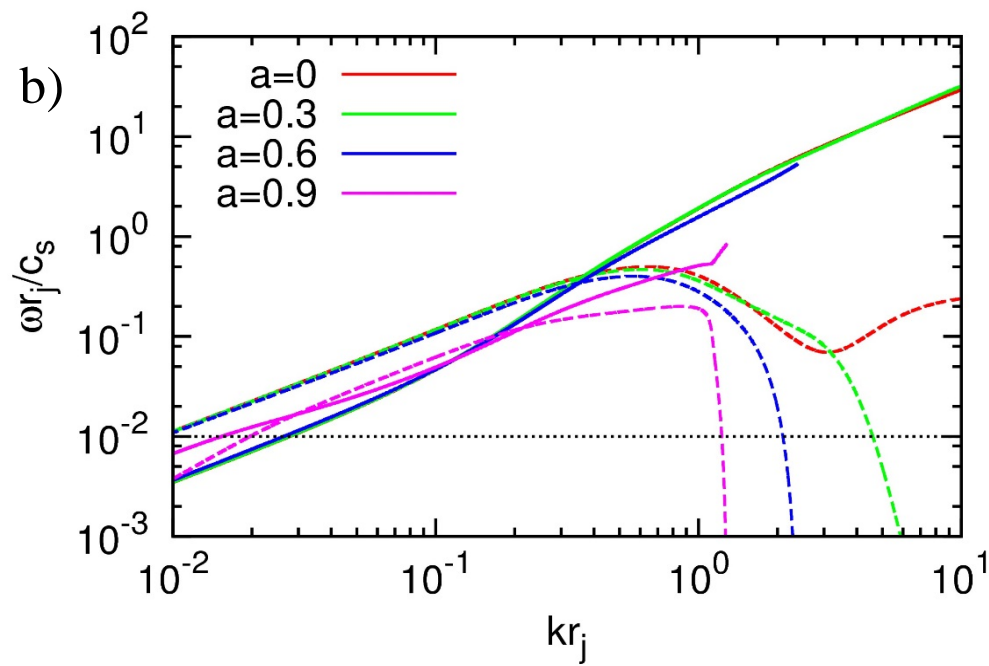

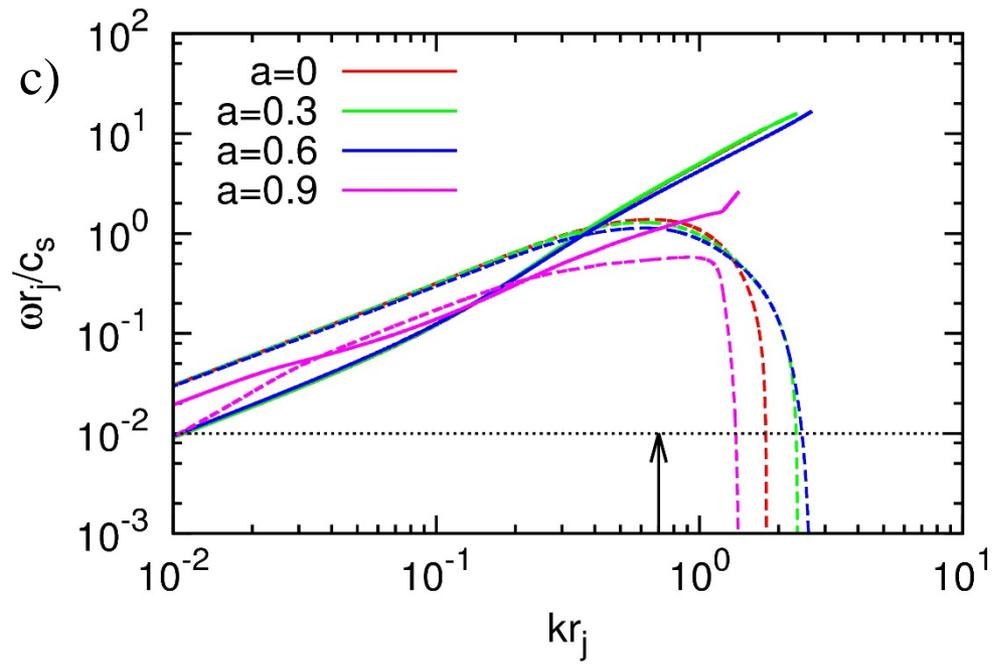

Fig. 9 shows the angular frequency (solid line) and temporal growth rate (dashed line) versus longitudinal wavenumber k for kink (m=1) fundamental mode of a non-magnetized jet. In Fig. 9a, the jet has M=4 and η=0.1. Increasing values of the parameter "a" indicate increasing shear, with a=0 (no shear) to a=0.9 (maximal shear). Fig. 9b shows the same information for a jet with β=1 (i.e. magnetic field is in equipartition with gas pressure). Fig. 9c shows the same information for a jet with β=0.5 (magnetically dominated).

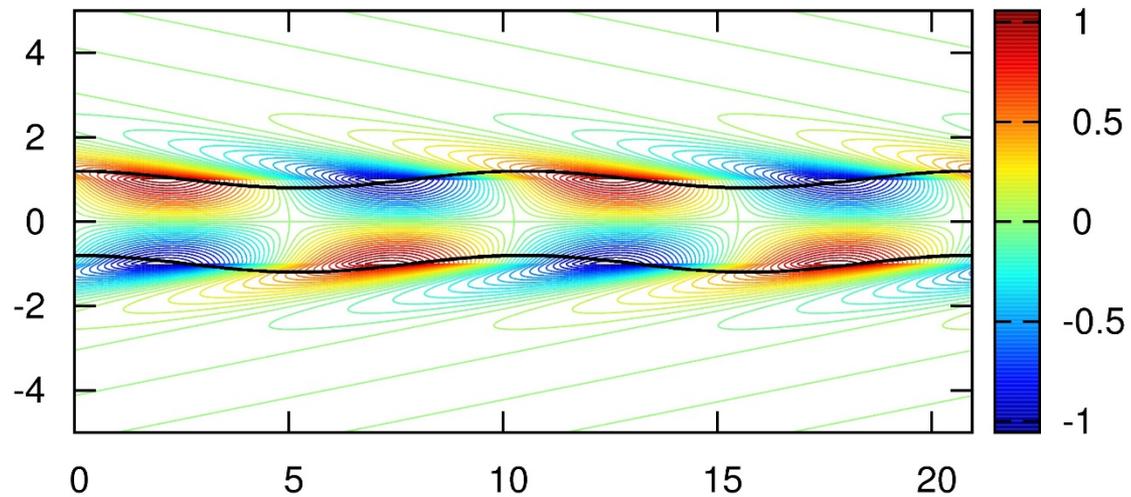

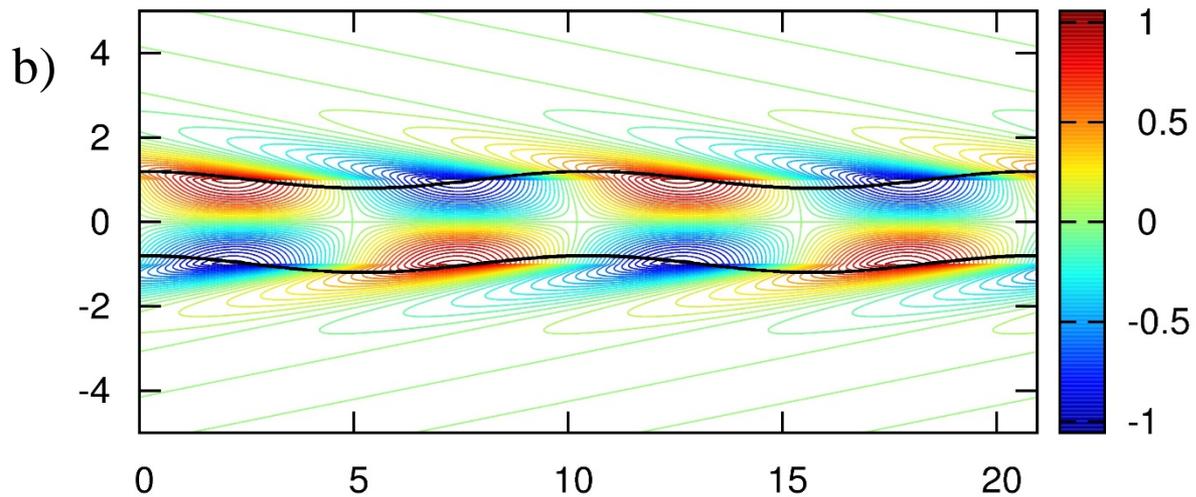

c) 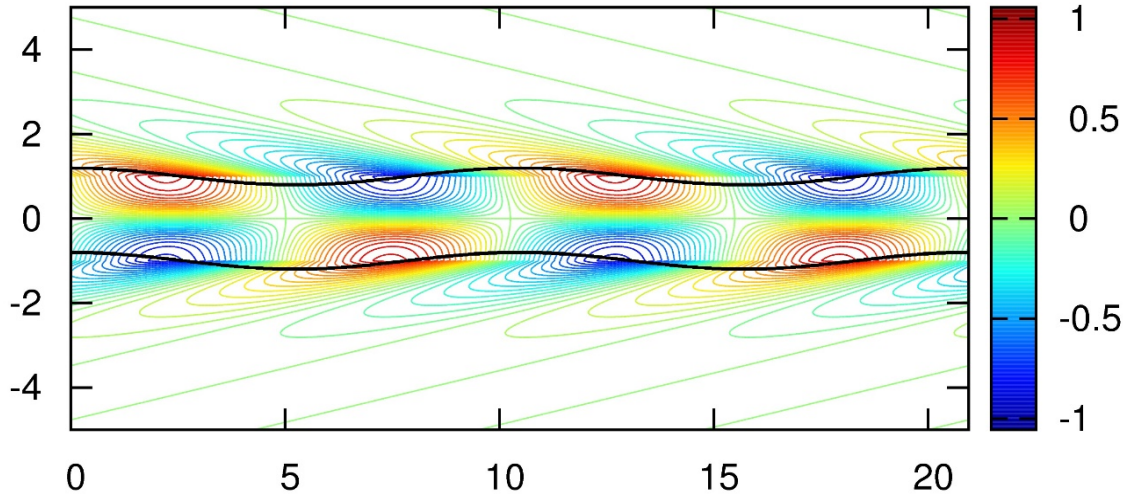

d) 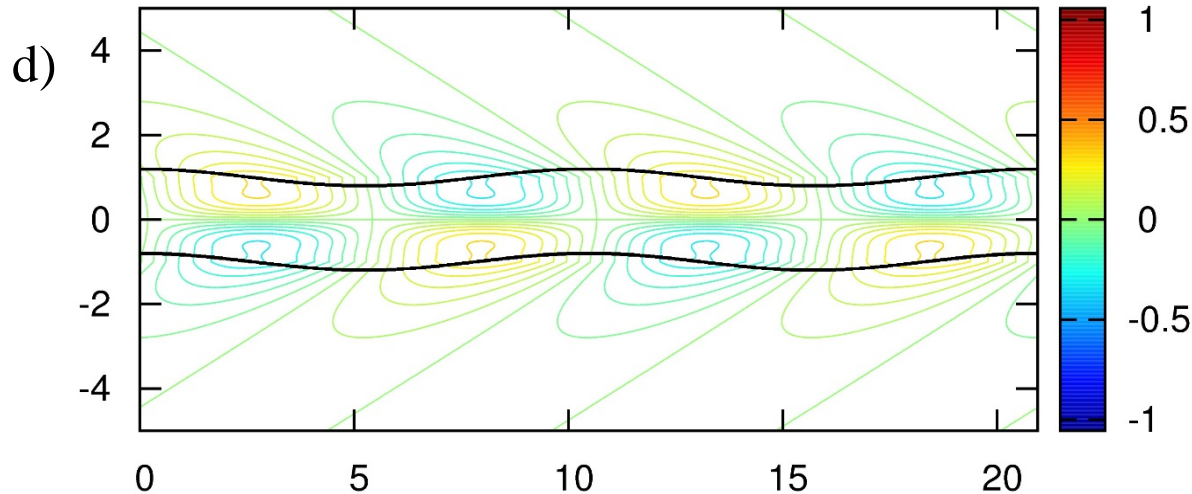

*Fig. 10a shows the pressure variation in a non-magnetized jet with a top hat profile. Fig. 10b shows the pressure variation in a non-magnetized jet with a=0.3 (mild shear). Fig. 10c shows the pressure variation in a non-magnetized jet with a =0.6 (modest shear). Fig. 10d shows the pressure variation in a non-magnetized jet with a=0.9 (strong shear). For all the cases in Fig. 10 we have k $r_j$ = 0.6. In all the cases, the jet's boundary has a fluctuation that is 20% of the jet's radius. The pressures are all on the same scale so that the pressures across panels within a figure can be inter-compared.*

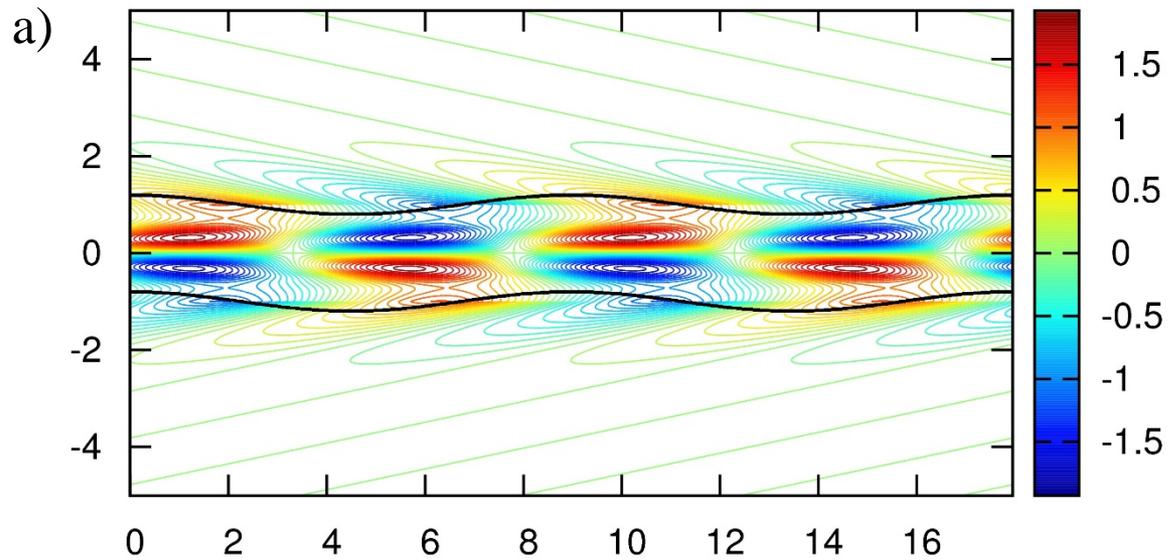
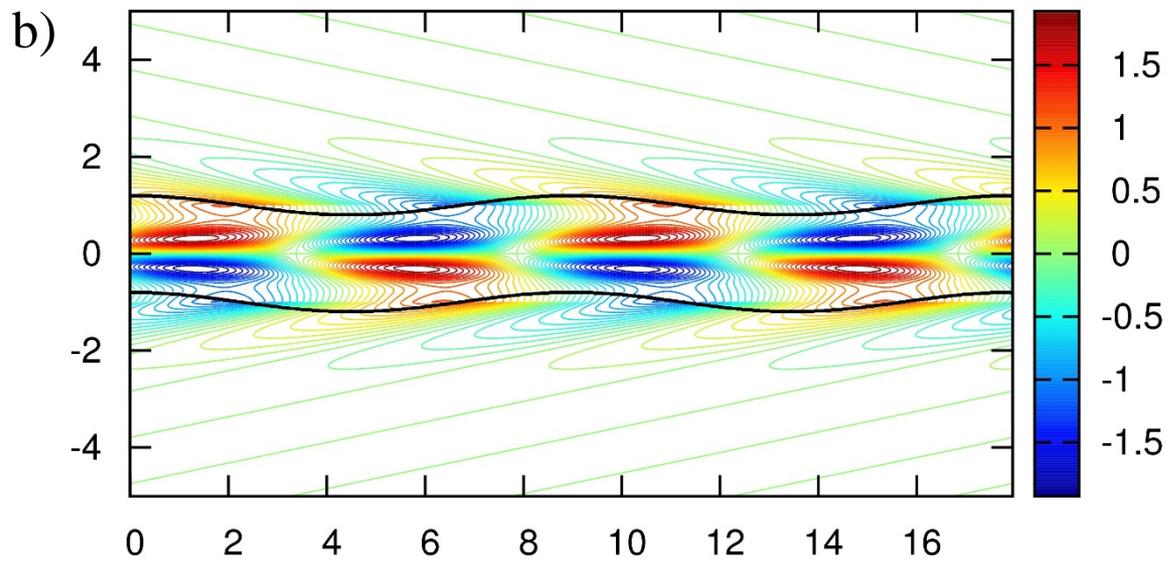

c)

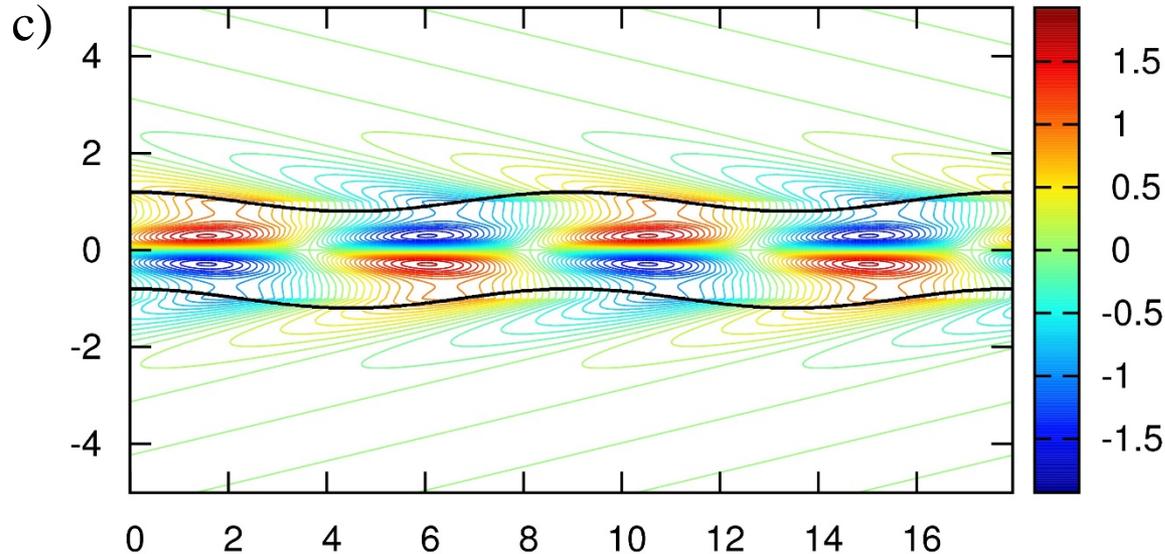

d)

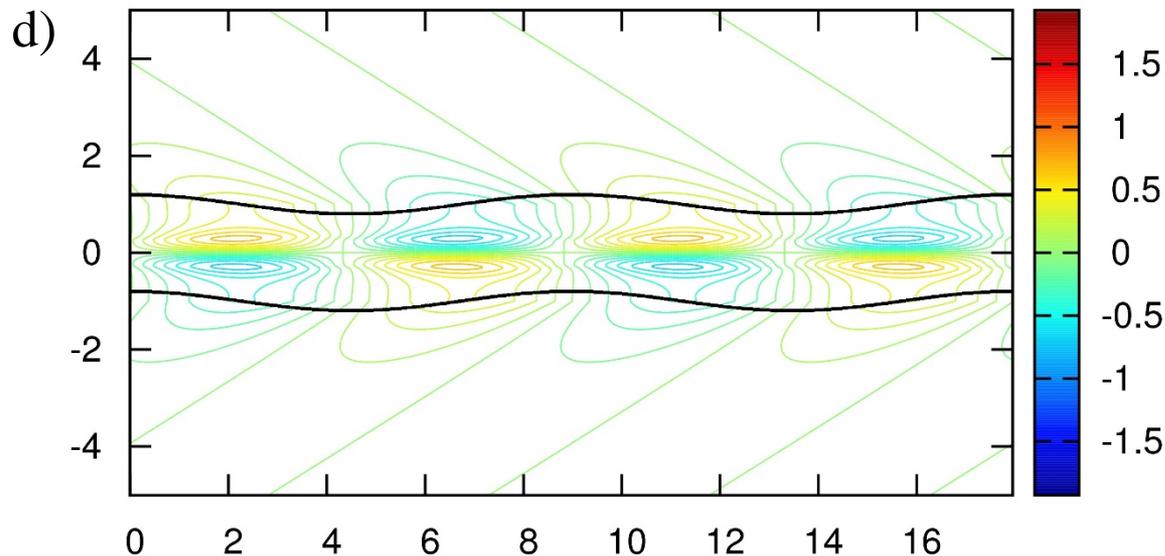

*Fig. 11a shows the pressure variation in a strongly-magnetized jet ($\beta=0.5$) with a top hat profile. Fig. 11b shows the pressure variation in a strongly-magnetized jet with a=0.3 (mild shear). Fig. 11c shows the pressure variation in a strongly-magnetized jet with a =0.6 (modest shear). Fig. 11d shows the pressure variation in a strongly-magnetized jet with a=0.9 (strong shear). For all the cases in Fig. 11 we have $k\,r_j$ = 0.7. In all the cases, the jet's boundary has a fluctuation that is 20% of the jet's radius. The pressures are all on the same scale so that the pressures across panels within a figure can be inter-compared.*

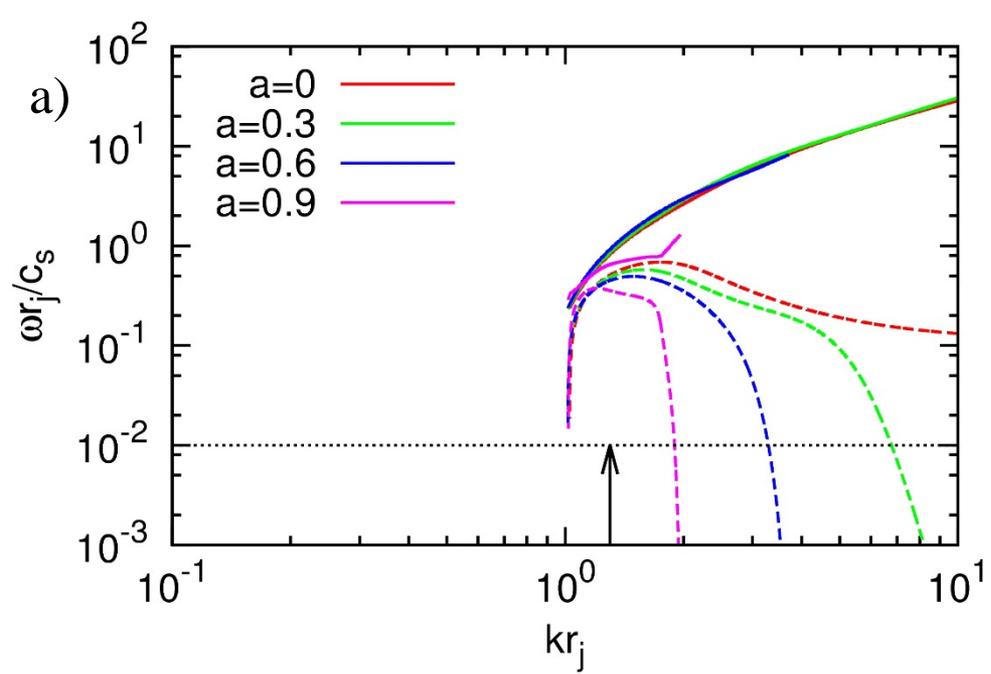

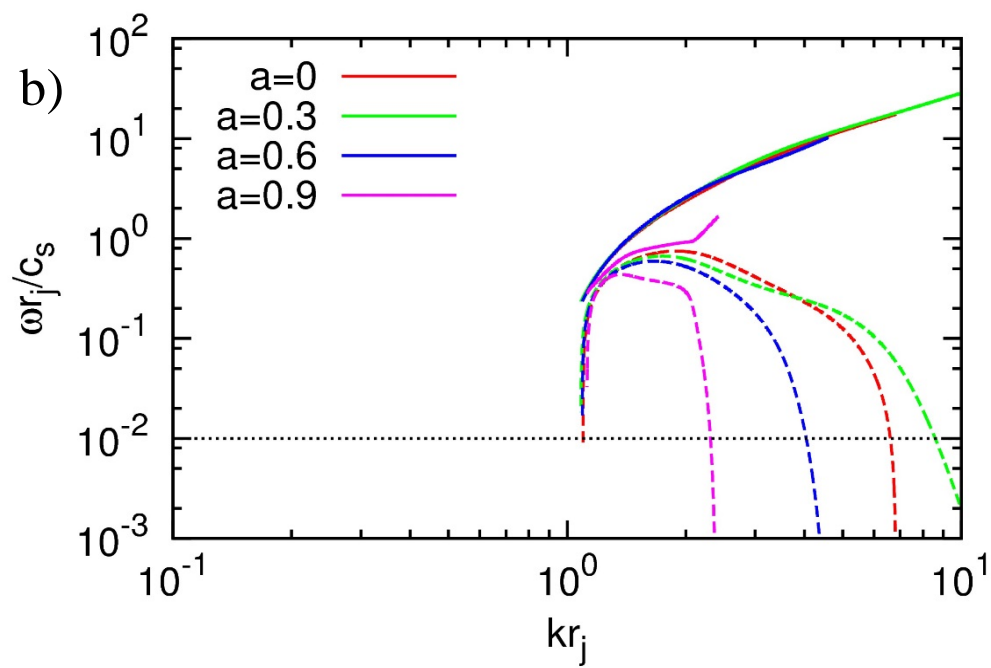

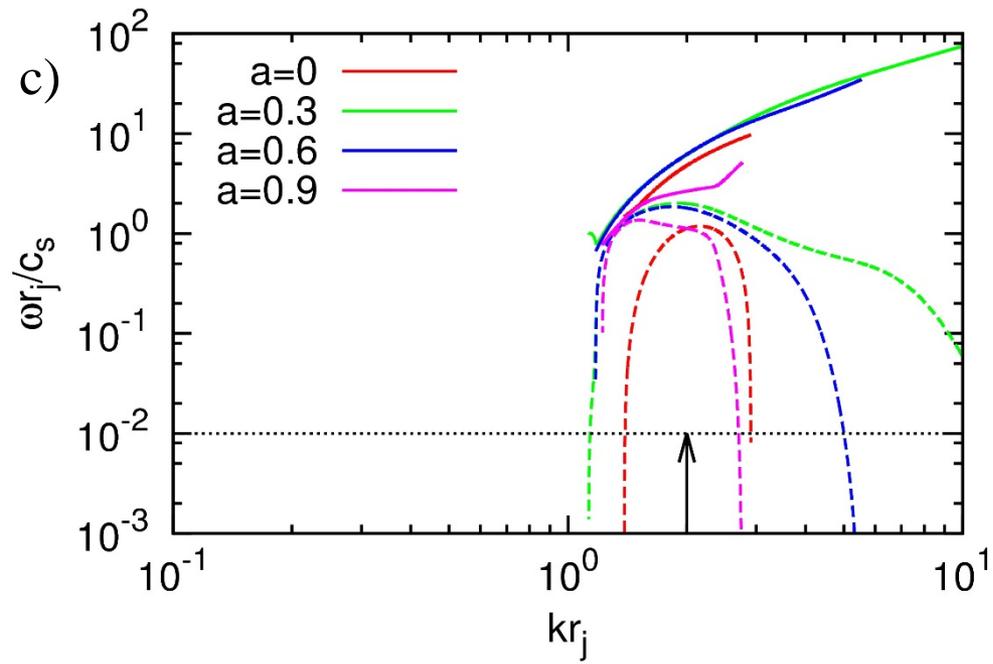

Fig. 12 shows the angular frequency (solid line) and temporal growth rate (dashed line) versus longitudinal wavenumber k for kink (m=1) 1st reflection mode of a non-magnetized jet. In Fig. 12a, the jet has M=4 and η=0.1. Increasing values of the parameter "a" indicate increasing shear, with a=0 (no shear) to a=0.9 (maximal shear). Fig. 12b shows the same information for a jet with β=1 (i.e. magnetic field is in equipartition with gas pressure). Fig. 6c shows the same information for a jet with β=0.5 (magnetically dominated).

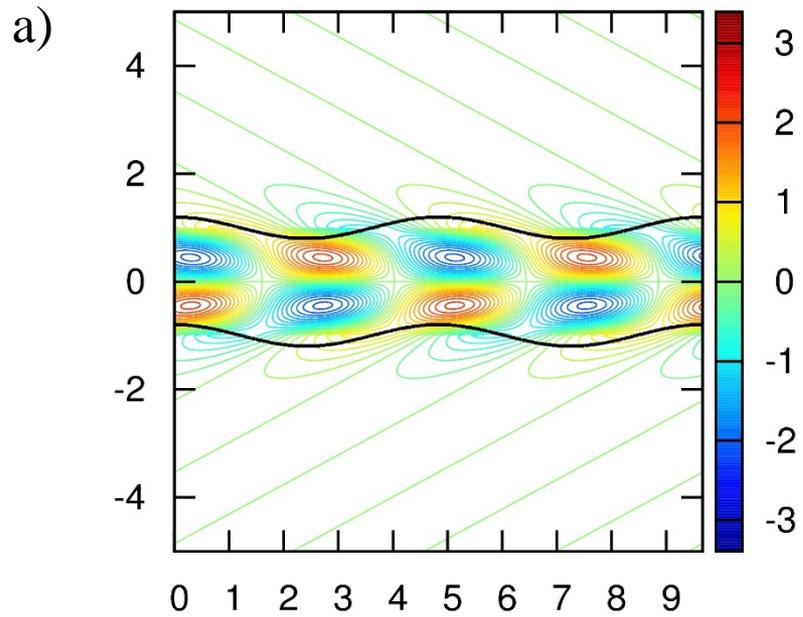

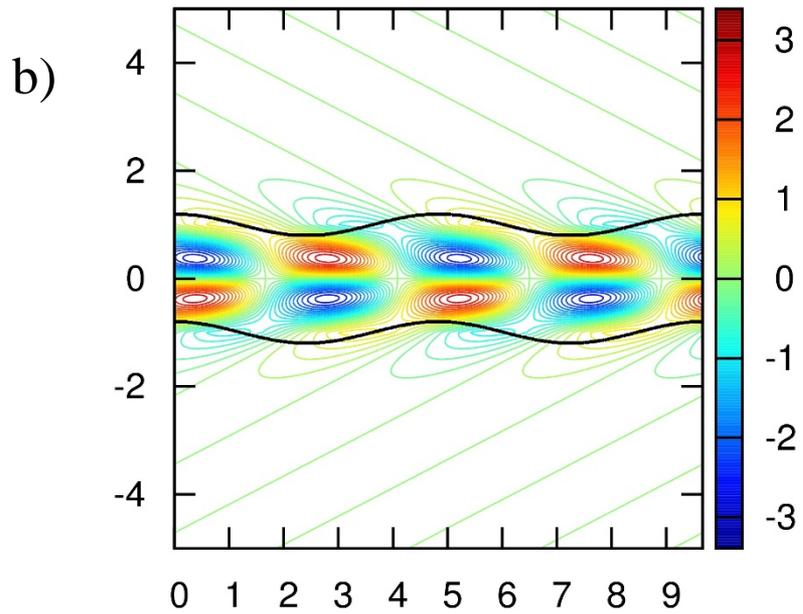

c) 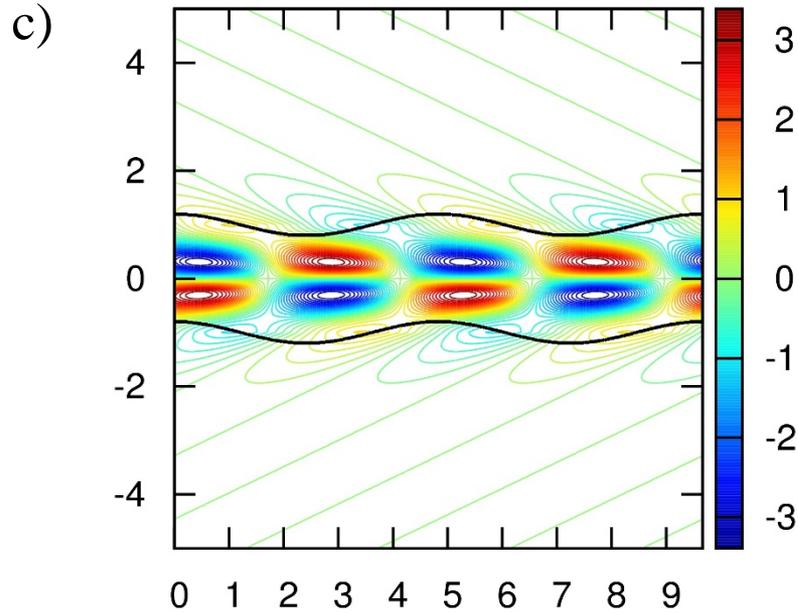

d) 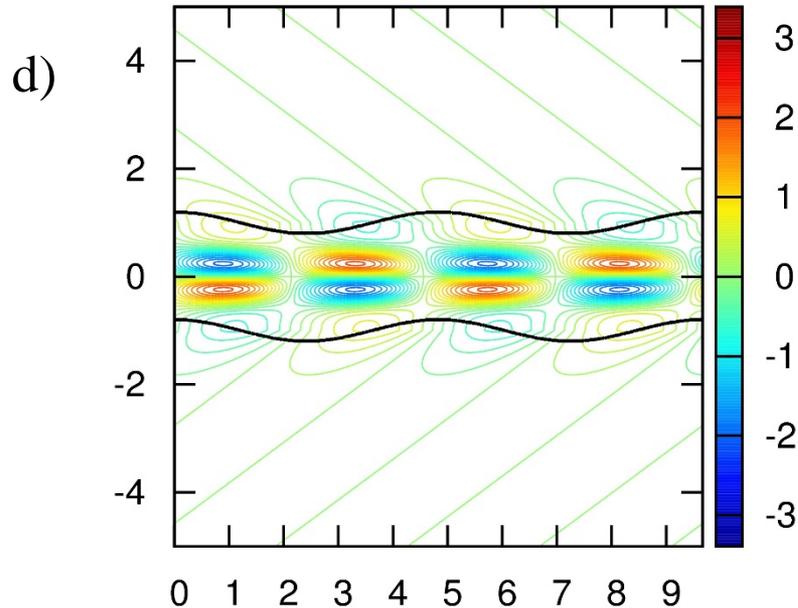

*Fig. 13a shows the pressure variation in a non-magnetized jet with a top hat profile. Fig. 13b shows the pressure variation in a non-magnetized jet with a=0.3 (mild shear). Fig. 13c shows the pressure variation in a non-magnetized jet with a =0.6 (modest shear). Fig. 13d shows the pressure variation in a non-magnetized jet with a=0.9 (strong shear). For all the cases in Fig. 13 we have k $r_j$ = 1.3. In all the cases, the jet's boundary has a fluctuation that is 20% of the jet's radius. The pressures are all on the same scale so that the pressures across panels within a figure can be inter-compared.*

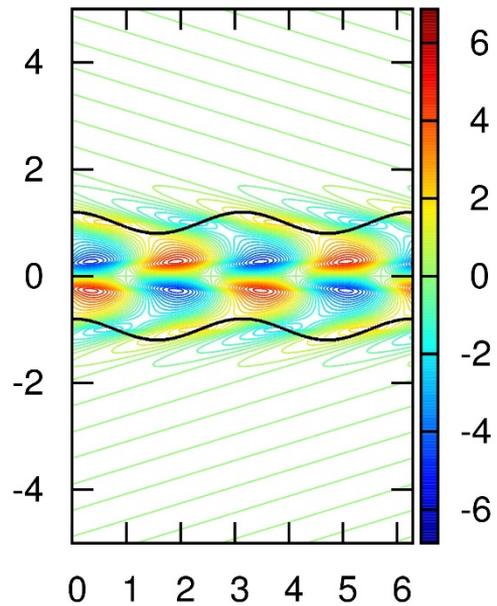

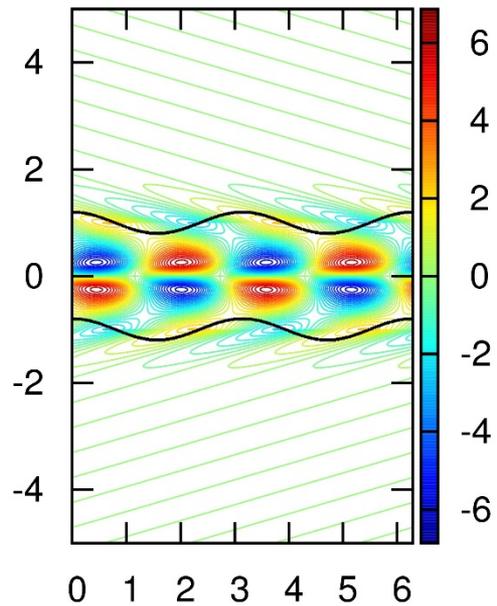

c) 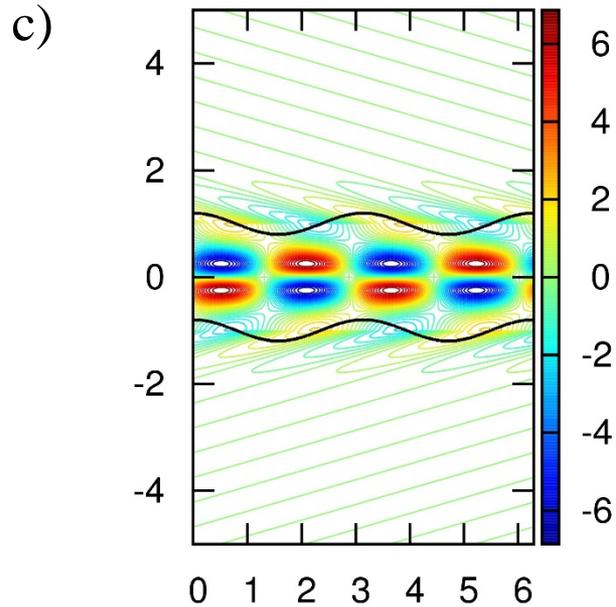

d) 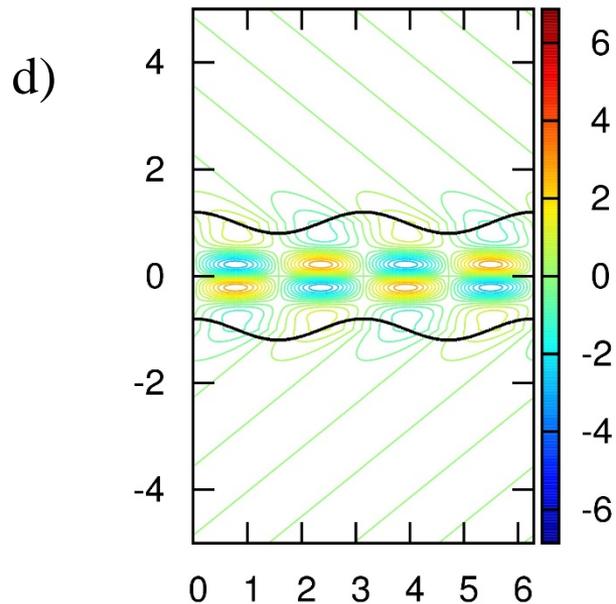

*Fig. 14a shows the pressure variation in a strongly-magnetized jet (β=0.5) with a top hat profile. Fig. 14b shows the pressure variation in a strongly-magnetized jet with a=0.3 (mild shear). Fig. 14c shows the pressure variation in a strongly-magnetized jet with a =0.6 (modest shear). Fig. 14d shows the pressure variation in a strongly-magnetized jet with a=0.9 (strong shear). For all the cases in Fig. 14 we have k $r_j$ = 2. In all the cases, the jet's boundary has a fluctuation that is 20% of the jet's radius. The pressures are all on the same scale so that the pressures across panels within a figure can be inter-compared.*